\documentclass[journal,10pt]{IEEEtran}

\usepackage{amsfonts}
\usepackage{graphicx}
\usepackage{amsmath}
\usepackage{amssymb}
\usepackage[caption=false]{subfig}
\usepackage[noadjust]{cite}
\usepackage{float}
\usepackage{algorithm}
\usepackage{bibentry}
\usepackage{balance}
\usepackage{algorithm}
\usepackage{algorithmicx}
\usepackage{algpseudocode}
\usepackage{xcolor}
\usepackage{balance}
\usepackage{multirow}
\usepackage{url}
\usepackage{cases}

\newcommand*\dd{\mathop{}\!\mathrm{d}}

\newtheorem{theorem}{Theorem}
\newtheorem{remark}{Remark}
\newtheorem{corollary}{Corollary}

\DeclareMathOperator*{\argmax}{argmax}
\DeclareMathOperator*{\argmin}{argmin}
\graphicspath{{img/}}

\begin{document}

\title{Physical Layer Security for NOMA Transmission in mmWave Drone Networks}
\author{Yavuz Yap{\i}c{\i}, Nadisanka Rupasinghe, \.{I}smail G\"{u}ven\c{c}, Huaiyu Dai and Arupjyoti Bhuyan
\thanks{Y. Yap{\i}c{\i} is with the Department of Electrical Engineering, University of South Carolina, Columbia, SC, 29201 (e-mail: yyapici@mailbox.sc.edu).}
\thanks{\.{I}. G\"{u}ven\c{c} and H. Dai are with the Department of Electrical and Computer Engineering, North Carolina State University, Raleigh, NC, 27606 (e-mail:~\{yyapici, iguvenc, hdai\}@ncsu.edu).}
\thanks{N. Rupasinghe is with DOCOMO Innovations, Inc., Palo Alto, CA, 94304  (e-mail: nrupasinghe@docomoinnovations.com).}
\thanks{A. Bhuyan is with Idaho National Labs, Idaho Falls, ID (e-mail: arupjyoti.bhuyan@inl.gov).}
\thanks{Work supported through the INL Laboratory Directed Research \& Development (LDRD) Program under DOE Idaho Operations Office Contract DE-AC07-05ID14517.}
}%

\maketitle

\begin{abstract}
The non-orthogonal multiple access (NOMA) and millimeter-wave (mmWave) transmission enable the unmanned aerial vehicle (UAV) assisted wireless networks to provide broadband connectivity over densely packed urban areas. The presence of malicious receivers, however, compromise the security of the UAV-to-ground communications links, thereby degrading secrecy rates. In this work, we consider a NOMA-based transmission strategy in a mmWave UAV-assisted wireless network, and investigate the respective secrecy-rate performance rigorously. In particular, we propose a \textit{protected-zone approach} to enhance the secrecy-rate performance by preventing the most \textit{vulnerable} subregion (outside the user region) from the presence of malicious receivers. The respective secrecy rates are then derived analytically as a function of the particular protected zone, which verifies great secrecy rate improvements through optimizing shape of the protected zone in use. Furthermore, we show that the optimal protected zone shape for mmWave links appears as a compromise between protecting the \textit{angle} versus \textit{distance} dimension, which would otherwise form to protect solely the distance dimension for sub-6GHz links. We also numerically evaluate the impact of transmission power, protected-zone size, and UAV altitude on the secrecy-rate performance improvement for the sake of practical deployments. 
\end{abstract}


\section{Introduction} \label{sec:intro}

The recent years have witnessed tremendous increase in mobile data traffic, which is estimated to reach $77.5$ exabytes per month by $2022$ with a $46\%$ growth rate. Moreover, the number of mobile devices is predicted to exceed $12.3$ billion by $2022$ with more than $422$ million being 5G capable \cite{Cisco2019visnet}. Towards supporting this enormous amount of data traffic and mobile users, the non-orthogonal multiple access (NOMA) and millimeter-wave (mmWave) communications are envisioned as two key technologies for the next-generation cellular networks \cite{Xia2019mmW}. Moreover, integrating unmanned aerial vehicles (UAV) into the existing terrestrial wireless networks is yet another powerful direction to seamlessly provide broadband connectivity over densely packed urban environments \cite{Zhang2019uavCom}. Along with the standardization efforts for 5G and beyond, this particular research direction connecting mmWave NOMA transmission and air-to-ground UAV links has been studied extensively in the literature from various perspectives over the past several years.

The confidentiality of the communications messages is yet another important challenge for the next-generation cellular networks. In general, the presence of illegitimate malicious receivers (i.e., eavesdroppers, or, in short, Eves) compromise the security of the information exchange between the transmit and receive ends. The physical layer security (PLS) has emerged as an effective solution to such privacy concerns in wireless connections \cite{Arslan2019PHYSECSur}. The protected zone and artificial noise are two well-known PLS techniques in the literature. The goal of both techniques is to degrade the communications links between the transmitter and Eves either through preventing the presence of Eves in specified regions (i.e., protected zone) \cite{Swami2013PHYLay, Quek2014EnhSec, Ding2016OnErg}, or by intentionally putting noise towards Eves (i.e., artificial noise) \cite{Negi2008GuaSec,Yin21017SecTra}. The protected zone requires additional measures to be taken on the ground (to clear the specified zone from the presence of Eves) while the artifical noise techniques are taken care of at the transmitter at the expense of increased complexity due to the special beamforming strategies.

Despite a solid body of studies considering information confidentiality for UAV networks (e.g., see \cite{Jiang2019UAV,Ng2019PhyLay,Zeng2019PhyLay,Zhang2020UAV} and references therein), there are limited number of works in the literature considering the PLS techniques for UAV-assisted communications in the mmWave frequency band \cite{Sun2019SecTra,Sun2019PhyLay,Sun2019SecCom,Cai2020SecmmW,Shi2019SecmmW}. In particular, \cite{Sun2019SecTra} considers a mmWave communications scenario involving an energy-constrained Internet-of-Things (IoT) receiver along with UAVs acting as relays. The respective secrecy rates are enhanced through optimizing the power allocation scheme and UAV location. In \cite{Sun2019PhyLay}, the authors consider a similar communications setting where the impact of small-scale fading is also considered during the secrecy-rate optimization. In a recent work, the authors incorporate NOMA into the PLS problem considered by \cite{Sun2019SecTra} and \cite{Sun2019PhyLay} with UAV links and mmWave transmission, which aims at evaluating the performance of a transceiver based on simultaneous wireless information and power transfer (SWIPT) \cite{Sun2019SecCom}. 

In addition, \cite{Cai2020SecmmW} investigates PLS enhancement methods for mmWave SWIPT UAV networks,and the directional modulation (DM) technique is adopted considering random frequency diverse array (RFDA) to meet security need. In \cite{Shi2019SecmmW}, the authors consider a mmWave communications scenario which is assisted by multiple UAVs, and a closed-form secrecy outage probability is derived in the presence of jammers and eavesdroppers. A cooperative jamming scheme is also proposed in which some of the UAVs transmit the jamming signal to degrade the quality of the eavesdropping channels. None of these works, however, investigate the protected-zone approach to evaluate the respective secrecy-rate performance improvement in UAV-assisted wireless networks in mmWave frequency band. Despite not considering mmWave UAV communications, the authors in \cite{Dai2018ReiLea} also consider the information confidentiality of the NOMA transmission against smart jammers being capable of adjusting their jamming strategy based on the communications setting, and propose a reinforcement-learning based strategy as a remedy.  

In this work, which is an extended version of \cite{Yapici2018EnhPhy}, we investigate the secrecy-rate performance of a UAV-assisted wireless communications scenario in mmWave frequency bands. In particular, we consider the NOMA transmission to improve the spectral efficiency through serving multiple legitimate users at the same time. The degradation in the secrecy rates due to the presence of surrounding Eves is mitigated through the formation (and optimization) of a \textit{protected zone} on the ground. The specific contributions of our work can be summarized as follows:
\begin{itemize}
    
    \item[---] We develop a complete framework where the secrecy-rate performance of the NOMA-based hybrid transmission scheme is rigorously investigated as a function of the shape and size of the protected zone. The respective outage and rate expressions are derived analytically, which show a very good match with the simulation data under various operation altitude and transmit power values, and verify the superior performance as compared to OMA.    
    
    \item[---] We rigorously investigate the impact of decoding capability at Eves on the secrecy-rate performance through \textit{best-case} (no multiuser decoding) and \textit{worst-case} (same multiuser decoding capability as legitimate users) conditions. We show that the upper bound provided by the best-case condition becomes very tight in the high transmit-power regime for sufficiently large protected zone (by approaching to that of worst-case condition) regardless of the particular power allocation ratio.
    
    \item[---] We also consider the impact of the particular Eve region sizes (in comparison to the protected zone area) on the secrecy-rate performance being \textit{space-} versus \textit{shape-limited}. We show that the performance is limited by the available space (i.e., space-limited) for small Eve regions or large protected zones. In contrast, the specific \textit{shape} of the protected zone dominates the secrecy-rate performance (i.e., shape-limited) for large Eve regions or small protected zones as a compromise between \textit{protecting} angle- versus distance-dimension, thanks to the mmWave propagation characteristics.

    \item[---] The minimum transmit power and size of the protected zone to achieve specific secrecy-rate performance are also investigated to keep the required ground and spectral resources as small as possible. The multiuser decoding capability at Eves and operation altitude are individually taken into consideration while evaluating the respective impact on these minimum values.   
    
\end{itemize}
    
The rest of the paper is organized as follows. Section~\ref{sec:system_model} introduces the system model under consideration. The NOMA transmission and secrecy rates are investigated in Section~\ref{sec:noma_transmission} with the respective outage and rate derivation in Section~\ref{sec:secrecy_sum_rates}. The numerical results are provided in Section~\ref{sec:results}, and the paper concludes with some remarks in Section~\ref{sec:conclusion}.

\section{System Model} \label{sec:system_model}

In this section, we first present the system model along with the mmWave channel and user distributions, and then describe the protected zone approach aiming at improving secrecy rates.

\subsection{Overview}

\begin{figure}[!t]
\begin{center}
\includegraphics[width=0.46\textwidth]{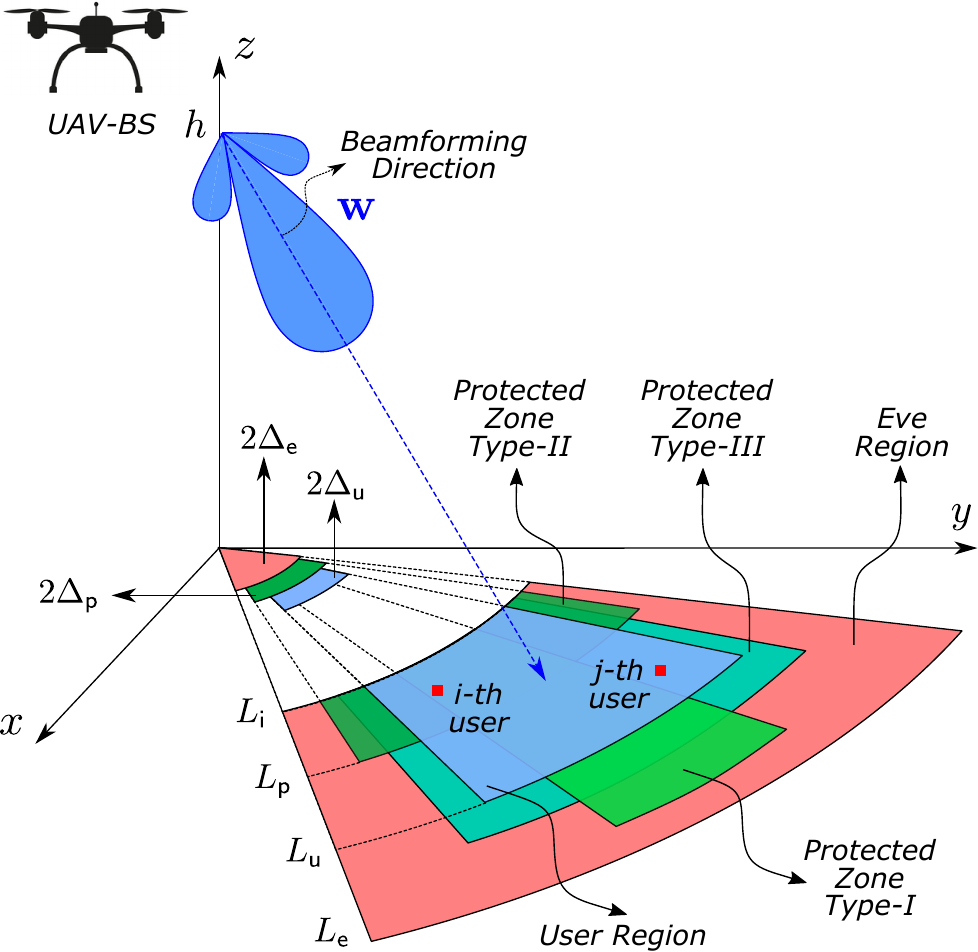}
\end{center}
\caption{System scenario where the NOMA transmission serves multiple users simultaneously in the presence of Eves.}
\label{fig:footprint}
\end{figure} 

We consider a NOMA transmission scenario where a stationary UAV base station (UAV-BS) equipped with an $M$~element antenna array is serving single-antenna users in mmWave downlink communications. We assume that all the users lie inside a specific \emph{user region} as shown in Fig.~\ref{fig:footprint}. A 3-dimensional (3D) beam is generated by the UAV-BS which entirely covers the user region. We assume that there are $K$ users in total, and the users are represented by the set $\mathcal{N}_\mathsf{U} \,{=}\, \{1,2,\ldots K\}$. The user region is specified by the inner radius $L_\mathsf{i}$, outer radius $L_\mathsf{u}$, and angle $\Delta_\mathsf{u}$, which is half of the fixed angle within the projection of horizontal propagation pattern of the UAV-BS on the $xy$--plane. Note that it is possible to reasonably model various different hot spot scenarios such as a stadium, concert hall, traffic jam, and urban canyon by modifying these control parameters.

We assume that although the user region is free from Eves, the surrounding region does include Eves trying to intercept the transmission between the UAV-BS and legitimate users. We specify a \textit{bounded region} around the user region, referred to as \textit{Eve region}, which is assumed to include all the Eves. Similar to the user region, we identify the Eve region by the inner radius $L_\mathsf{i}$ (same as the user region), outer radius $L_\mathsf{e}$ (such that $L_\mathsf{e} \,{\geq}\, L_\mathsf{u}$), and angle $\Delta_\mathsf{e}$ (such that $\Delta_\mathsf{e} \,{\geq}\, \Delta_\mathsf{u}$), as shown in Fig.~\ref{fig:footprint}. We assume $K_\mathsf{e}$ Eves in total, which are represented by the set $\mathcal{N}_\mathsf{E} \,{=}\, \{1,2,\ldots K_\mathsf{e}\}$. Note that horizontal footprint of the UAV-BS beam pattern covers the Eve region (so that any Eve has nonzero channel to the UAV-BS), as well, but the coverage over Eve region might be provided by the side lobes depending on the specific radiation pattern. 

Note that although we formulate the overall problem through a fixed and \textit{bounded} Eve region, we present numerical results in Section~\ref{sec:results} for various Eve-region sizes. To this end, we express the outer radius and angle of the Eve region in terms of the user region parameters as $L_\mathsf{e} \,{=}\, \kappa L_\mathsf{u}$ and $\Delta_\mathsf{e} \,{=}\, \kappa \Delta_\mathsf{u}$, respectively, where $\kappa$ is referred to as \textit{expansion ratio}. In particular, we consider large enough $\kappa$ values to investigate the impact of i) the relative Eve region size, and ii) the design of the shape of the protected zone on the overall secrecy rates.

\subsection{Location Distribution and mmWave Channel Model} \label{sec:location_dist} 

We assume no special location distribution for the legitimate users and Eves, and hence assume that they are uniformly distributed within the user and Eve region, respectively, following homogeneous Poisson point process (HPPP) with the densities $\lambda_\mathsf{u}$ and $\lambda_\mathsf{e}$, respectively. The number of users and Eves are therefore Poisson distributed such that $\mathsf{Pr}(k \textrm{ users in the user region}) \,{=}\, \mu_\mathsf{u}^k \, e^{{-}\mu_\mathsf{u}}/k!$ and $\mathsf{Pr}(k \textrm{ Eves in the Eve region}) \,{=}\, \mu_\mathsf{e}^k e^{{-} \, \mu_\mathsf{e}}/k!$ with the mean densities $\mu_\mathsf{u} \,{=}\, \mathsf{A}_\mathsf{u} \lambda_\mathsf{u}$ and $\mu_\mathsf{e} \,{=}\, \mathsf{A}_\mathsf{e} \lambda_\mathsf{e}$, where area of the user and Eve regions are given as $\mathsf{A}_\mathsf{u} \,{=}\, \left(L_\mathsf{u}^2\,{-}\,L_\mathsf{i}^2\right) \Delta_\mathsf{u}$ and $\mathsf{A}_\mathsf{e} \,{=}\, \left( L_\mathsf{e}^2 \,{-}\, L_\mathsf{i}^2\right) \Delta_\mathsf{e} \,{-}\, \mathsf{A}_\mathsf{u}$, respectively.

We assume that all the users have line-of-sight (LoS) paths since i) LoS path is much stronger than the non-LoS (NLoS) paths in mmWave frequencies~\cite{Ding17PoorRandBeamforming}, and ii) multipath scattering is assumed to be not significant thanks to the nature of our scenario (e.g., UAV is positioned with a clear LoS view towards users). The channel between the UAV-BS and $k$-th user is therefore given as
\begin{align} \label{eq:k_UE_original_channel}
\textbf{h}_k = \sqrt{M} \frac{\alpha_k \, \textbf{a}(\theta_k)}{\left[\mathsf{PL}\left(\sqrt{d_k^2 + h^2}\right)\right]^{\frac12}},
\end{align} 
where $h$, $d_k$, $\alpha_{k}$ and $\theta_{k}$ represent the UAV-BS altitude, horizontal distance between the $k$-th user and UAV-BS, small-scale fading gain being standard complex Gaussian, and angle-of-departure (AoD) projected onto the $xy$--plane, respectively. In addition, $\mathsf{PL}\left(x\right)$ in \eqref{eq:k_UE_original_channel} stands for the frequency-dependent path loss (PL) over the LoS distance of $x$ (to be detailed in Section~\ref{sec:results}), and $\textbf{a}(\theta_{k})$ is the steering vector associated with $\theta_{k}$, for which the $m$-th element is given as
\begin{align}
\Big[ \textbf{a}(\theta_{k}) \Big]_m =  \frac{1}{\sqrt{M}} {\rm exp} \left\lbrace {-}j2\pi \frac{d_\mathsf{s}}{\lambda_\mathsf{c}} \left( m{-}1\right) \cos\left( \theta_{k} \right) \right\rbrace,
\end{align}
for $m \,{=}\, 1,\dots,M$, where $d_\mathsf{s}$ is the antenna element spacing of uniform linear array (ULA) under consideration, and $\lambda_\mathsf{c}$ is the wavelength of the carrier frequency. Note that channel of any Eve in the Eve region and UAV-BS can also be described by \eqref{eq:k_UE_original_channel} using the appropriate parameters for the Eves (i.e., horizontal distance, small-scale fading, and AoD).

\subsection{Protected Zone Approach} \label{subsec:protected_zone}

The overall transmission scheme between the UAV-BS and legitimate users is highly prone to eavesdropping, and the confidentiality of the links is accordingly compromised. In this study, we consider \textit{protected zone} approach as an attempt to enhance the secrecy rates through PLS measures~\cite{Swami2013PHYLay, Hanzo2017EnhPhy}. In the proposed approach, an additional area, referred to as protected zone, around the user region (and inside the Eve region) has been perfectly cleared from Eves by taking necessary physical measures on the ground. Since ensuring no Eves within the protected zone requires certain resources being spent on the ground, \textit{the area of this protected zone is desired to be kept as small as possible for the given secrecy-rate goal}. Equivalently, enhancing secrecy rates as much as possible for a given protected zone area is yet another choice of the design goal, which we formulate in the subsequent sections. Towards this end, we show in Section~\ref{sec:noma_transmission} that the secrecy rates can be further improved by optimizing the shape of the protected zone by taking into account the propagation characteristics of the mmWave transmission.    

\begin{figure}[!t]
\begin{center}
\includegraphics[width=0.46\textwidth]{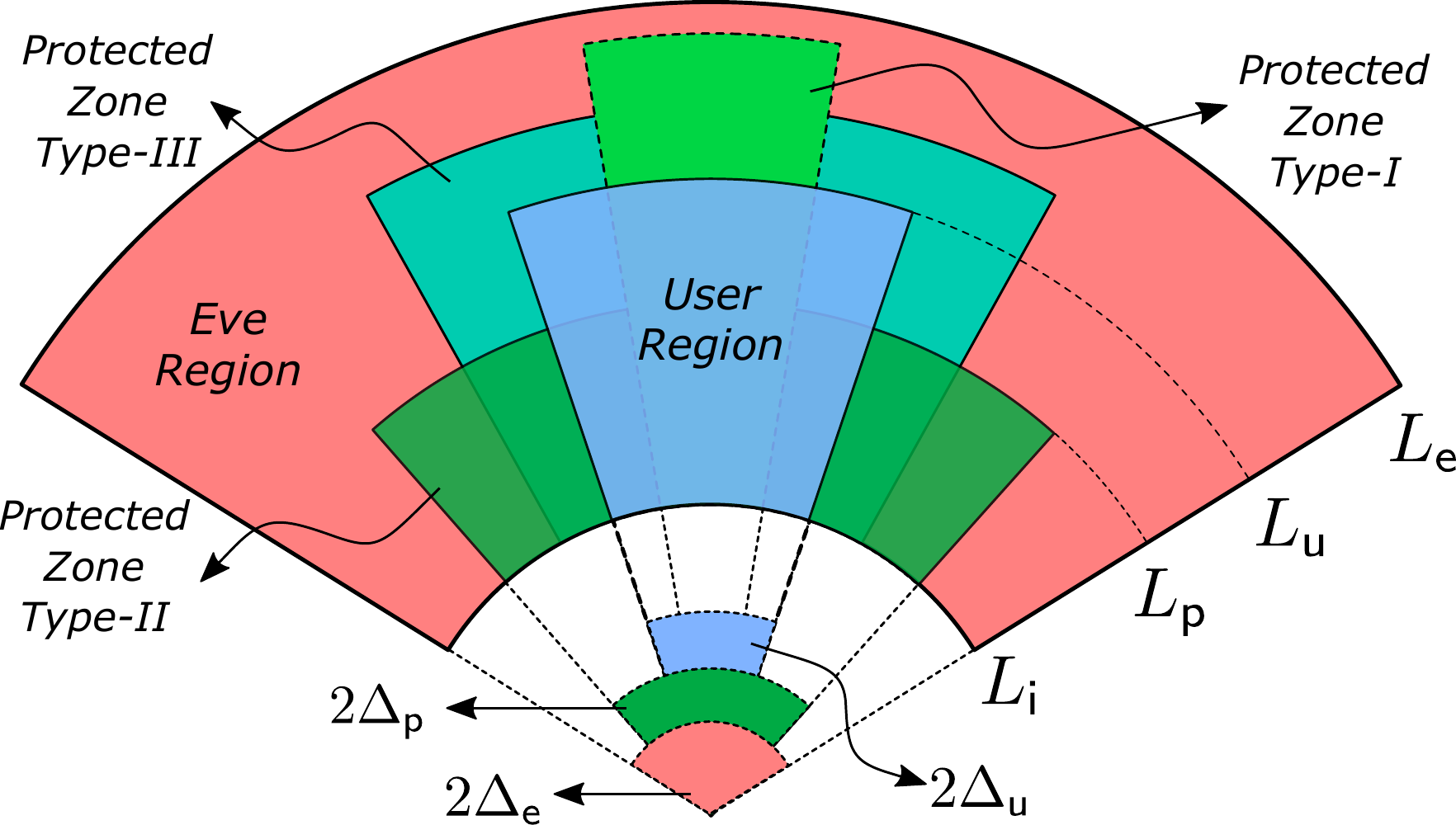}
\end{center}
\caption{Footprint of the user and Eve regions explicitly showing \textit{protected zone} 
represented by angle-distance pair $(\Delta_\mathsf{p},\, L_\mathsf{p})$ with example layouts of Type-I, Type-II, and Type-III.}
\label{fig:protected_zone}
\end{figure}

We describe the desired protected zone by an angle-distance (radius) pair $(\Delta_\mathsf{p},\, L_\mathsf{p})$ such that $\Delta_\mathsf{p} \,{\leq}\, \Delta_\mathsf{e}$ and $L_\mathsf{i} \,{\leq}\, L_\mathsf{p} \,{\leq}\, L_\mathsf{e}$, as shown in Fig.~\ref{fig:protected_zone}. The protected area is therefore a fraction of the complete Eve region, and we denote this area fraction by $q \,{=}\, \mathsf{A}_\mathsf{P}/\mathsf{A}_\mathsf{e} \,{\leq}\, 1$, where $\mathsf{A}_\mathsf{P}$ is the area of the protected region given as
\begin{subnumcases}{ \hspace{-0.3in} \mathsf{A}_\mathsf{P} \,{=} }
    \!\! \left(L_\mathsf{p}^2\,{-}\,L_\mathsf{u}^2 \right) \Delta_\mathsf{p}, & \hspace{-0.2in} if $L_\mathsf{u} \leq L_\mathsf{p} \leq L_\mathsf{e}$, \nonumber \\
    & $\Delta_\mathsf{p} \leq \Delta_\mathsf{u} $ \!, \label{eq:area_pzone_I}\\
    \!\! \left(L_\mathsf{p}^2\,{-}\,L_\mathsf{i}^2 \right)\left(\Delta_\mathsf{p}{-}\Delta_\mathsf{u} \right), & \hspace{-0.2in} if $L_\mathsf{i} \leq L_\mathsf{p} \leq L_\mathsf{u}$, \nonumber \\
    & $\Delta_\mathsf{u} \leq \Delta_\mathsf{p} \leq \Delta_\mathsf{e}$ \!, \label{eq:area_pzone_II}\\
    \!\! \left(L_\mathsf{p}^2\,{-}\,L_\mathsf{i}^2 \right) \Delta_\mathsf{p} \,{-}\, \mathsf{A}_\mathsf{u}, & \hspace{-0.2in} if $L_\mathsf{u} \leq L_\mathsf{p} \leq L_\mathsf{e}$, \nonumber \\
    & $\Delta_\mathsf{u} \leq \Delta_\mathsf{p} \leq \Delta_\mathsf{e}$ \!\!\! . \label{eq:area_pzone_III}
\end{subnumcases}

The protected zone can take different shapes for a fixed $q$ value (i.e., without changing its area $\mathsf{A}_\mathsf{P}$), as sketched in Fig.~\ref{fig:protected_zone}. In particular, we refer to the protected zones satisfying \eqref{eq:area_pzone_I}, \eqref{eq:area_pzone_II}, and \eqref{eq:area_pzone_III} as Type-I, Type-II, and Type-III, respectively, as shown in Fig.~\ref{fig:protected_zone}.  Note that whenever we have $\Delta_\mathsf{p} \,{\leq}\, \Delta_\mathsf{u}$, $L_\mathsf{p}$ should be sufficiently greater than $L_\mathsf{u}$ to have a nonzero protected zone (i.e. Type-I). On the other hand, when we have $\Delta_\mathsf{u} \,{\leq}\, \Delta_\mathsf{p} \,{\leq}\, \Delta_\mathsf{e}$, $L_\mathsf{p}$ might be smaller or greater than $L_\mathsf{u}$ (i.e., Type-II and Type-III, respectively) depending on the area of the user region and particular $q$ choice. 

Considering the algebraic definition of the different protected zone types in \eqref{eq:area_pzone_I}-\eqref{eq:area_pzone_III}, we can express $L_\mathsf{p}$ parametrically through geometrical relations in Fig.~\ref{fig:protected_zone} as follows
\begin{align}
L_\mathsf{p} = \begin{cases}
    \displaystyle \bigg[ L_\mathsf{u}^2 + \frac{ q\mathsf{A}_\mathsf{e} }{ \Delta_\mathsf{p} }  \bigg]^{\frac{1}{2}} \!\! , & \text{if Type-I} , \\
    \displaystyle \bigg[ L_\mathsf{i}^2 + \frac{ q\mathsf{A}_\mathsf{e} }{ \Delta_\mathsf{p}-\Delta_\mathsf{u} }  \bigg]^{\frac{1}{2}} \!\!, & \text{if Type-II} , \\
    \displaystyle
    \bigg[ L_\mathsf{i}^2 + \frac{ q\mathsf{A}_\mathsf{e} + \mathsf{A}_\mathsf{u} }{ \Delta_\mathsf{p} } \bigg]^{\frac{1}{2}} \!\!, & \text{if Type-III} .
\end{cases}
\end{align}
Note that depending on the expansion ratio $\kappa$, the minimum value that $L_\mathsf{p}$ can take (while meeting the area fraction $q$) is determined by the maximum angular width of the particular protected zone types (i.e., $\Delta_\mathsf{u}$ for Type-I, and $\kappa \Delta_\mathsf{u}$ for Type-II and Type-III). As a result, the minimum value $L_\mathsf{p}$ is given as follows
\begin{align}
L_\mathsf{p}^\mathsf{min} = \begin{cases}
    \Big[ \left( 1 + q \left( \kappa^3 - 1 \right) \right) L_\mathsf{u}^2 & \\
    \hspace{0.7in} - q \left( \kappa - 1 \right) L_\mathsf{i}^2 \Big]^{\frac{1}{2}} \!\! , & \text{if Type-I}, \\
    \Big[ q \left( 1 + \kappa + \kappa^2 \right) L_\mathsf{u}^2 & \\
    \hspace{0.7in} + \left( 1 - q \right) L_\mathsf{i}^2 \Big]^{\frac{1}{2}} \!\!, & \text{if Type-II},\\
    \kappa^{{-}\frac12}\Big[ \left( 1 + q \left( \kappa^3 - 1 \right) \right) L_\mathsf{u}^2 & \\ 
    \hspace{0.4in} + \left( 1- q \right) \left( \kappa - 1 \right) L_\mathsf{i}^2 \Big]^{\frac{1}{2}} \!\!, & \text{if Type-III}.
\end{cases}
\end{align}
In a similar fashion, the minimum value that $\Delta_\mathsf{p}$ can effectively take is determined by the maximum range (distance) of the protected zone given, which is given for each protected zone type as follows 
\begin{align*}
\Delta_\mathsf{p}^\mathsf{min} {=}\, \begin{cases}
    \displaystyle \frac{q}{(\kappa+1)L_\mathsf{u}^2} \bigg[ \left( 1 + \kappa + \kappa^2 \right) L_\mathsf{u}^2 - L_\mathsf{i}^2 \bigg] \Delta_\mathsf{u} , & \hspace{-0.08in} \text{if Type-I} , \vspace{0.1in}\\
    \displaystyle \bigg[ 1 + q \frac{ \left( \kappa^3 - 1 \right) L_\mathsf{u}^2 - \left( \kappa - 1 \right) L_\mathsf{i}^2 }{ L_\mathsf{u}^2 - L_\mathsf{i}^2 } \bigg] \Delta_\mathsf{u} , & \hspace{-0.08in} \text{if Type-II} , \vspace{0.1in} \\
    \displaystyle \bigg[ 1 + q \frac{ \left( \kappa^3 - 1 + \left( 1-\kappa^2 \right) / q \right) L_\mathsf{u}^2 }{ \kappa^2 L_\mathsf{u}^2 - L_\mathsf{i}^2 } \bigg] \Delta_\mathsf{u},  & \\
    \displaystyle \hspace{0.8in} - q \frac{ \left( \kappa - 1 \right) L_\mathsf{i}^2 }{ \kappa^2 L_\mathsf{u}^2 - L_\mathsf{i}^2 } \Delta_\mathsf{u},  & \hspace{-0.15in} \text{if Type-III}.
\end{cases}
\end{align*}
We list the common notations that we use throughout the paper in Table~\ref{tab:notations} for ease of readability.
\begin{table}[!t]
    \renewcommand{\arraystretch}{1.0}
	\caption{Symbols and Notations}
	\label{tab:notations}
	\centering
	\begin{tabular}{ll}
	\hline
	Parameter & Value \\
	\hline\hline
    $M$ & \# of UAV-BS antennas \\
    $K$ & \# of legitimate users \\
    $K_\mathsf{e}$ & \# of Eves \\ 
    $K_\mathsf{min}^\mathsf{N}$ & min \# of users for NOMA \\
    $K_\mathsf{min}^\mathsf{S}$ & min \# of users for SUT \\
    $\mathcal{N}_\mathsf{U}$ & set of all legitimate users \\
    $\mathcal{N}_\mathsf{N}$ & set of all NOMA users \\
    $\mathcal{N}_\mathsf{E}$ &	set of all Eves \\
    $\mathcal{S}_\mathsf{N}$ & set of \# of users for NOMA\\
    $\mathcal{S}_\mathsf{N}$ & set of \# of users for SUT\\
    $L_\mathsf{i}$ & inner radius of the user region \\
    $L_\mathsf{u}$ & outer radius of the user region \\
    $L_\mathsf{e}$ & outer radius of the Eve region \\
    $L_\mathsf{p}$ & radius of the protected \\
    $\Delta_\mathsf{u}$ & half-angle of the user region \\
    $\Delta_\mathsf{e}$	& half-angle of the Eve region \\
    $\Delta_\mathsf{p}$	& half-angle of the protected zone \\
    $\kappa$ & expansion ratio \\
    $\lambda_\mathsf{u}$ & user density \\
    $\lambda_\mathsf{e}$ & Eve density \\
    $\mu_\mathsf{u}$ & mean user density \\
    $\mu_\mathsf{e}$ & mean Eve density \\
    $\mathsf{A}_\mathsf{u}$	& user region area \\
    $\mathsf{A}_\mathsf{e}$	& Eve region area \\
    $\mathsf{A}_\mathsf{p}$	& protected zone area \\
    $\textbf{h}_k$ & user channel \\
    $h$ & UAV-BS altitude \\
    $d_k$ & horizontal distance \\
    $\alpha_{k}$ & small-scale fading \\
    $\theta_{k}$ & AoD \\
    $\overline{\theta}$	 &	beamforming direction \\
    $\mathsf{PL}$ & path loss \\
    $d_\mathsf{s}$ & antenna element spacing \\
    $\lambda_\mathsf{c}$ & carrier wavelength \\
    $q$	& area fraction \\
    $\beta_k$ & power allocation\\
    $\mathsf{P}_\mathsf{tx}$ & transmit power  \\
    $s_{k}$ & user message  \\
    $\textbf{w}$ & beamforming vector \\
    $N_0$ & noise variance \\
    $\mathsf{F}_M$ & Fej\'er kernel \\
    $\overline{\mathsf{R}}_k$ &	target secrecy rate \\
    $\Phi$ & given condition on $K$ and $K_\mathsf{e}$ \\
    \hline
	\end{tabular}
\end{table}

\section{Secure NOMA for UAV-BS Downlink} \label{sec:noma_transmission}

In this section, we consider the NOMA transmission in mmWave downlink between the UAV-BS and legitimate users, and evaluate the secrecy rates along with the protected zone optimization. 

\subsection{Transmission to Legitimate Users}

We assume that the desired legitimate users to be involved in the NOMA transmission are represented by the set $\mathcal{N}_\mathsf{N}$ such that $\mathcal{N}_\mathsf{N} \,{\subset}\, \mathcal{N}_\mathsf{U}$, and are indexed from the best to the worst with respect to their effective channel gains (to be defined specifically in the following). We define $\beta_k$ as the power allocation coefficient for the $k$-th user such that $\sum _{i\,{\in}\,\mathcal{N}_\mathsf{N}} \beta_i^2\,{=}\,1$. In order to yield a sufficient decoding performance, we assume that the NOMA users are allocated adequate power levels that are inversely proportional to channel qualities such that $\beta_i\,{<}\,\beta_j$ for $\forall \, i \,{<}\, j$ with $i,j \,{\in}\, \mathcal{N}_\mathsf{N}$. The transmitted signal is therefore generated by superposition coding as follows
\begin{align}\label{eq:superposition_code}
\textbf{x} = \sqrt{\mathsf{P}_\mathsf{tx}}\textbf{w}\sum \limits_{i\,{\in}\,\mathcal{N}_\mathsf{N}} \beta_i s_{i}, \end{align} 
where $\mathsf{P}_\mathsf{tx}$ and $s_{k}$ are the total downlink transmit power and $k$-th user's message, respectively, and $\textbf{w}$ is the beamforming vector of the UAV-BS such that its projection on the $xy$--plane is in the direction of an azimuth angle $\overline{\theta} \,{\in}\, [0{,}\,2\pi]$. The received signal at the $k$-th user is given by
\begin{align} \label{eq:k-th_user_Rx_signal}
y_{k}= \textbf{h}_{k}^{\rm H} \textbf{x} +  v_k = \sqrt{\mathsf{P}_\mathsf{tx}}\textbf{h}_{k}^{\rm H} \textbf{w}\sum \limits_{i\,{\in}\,\mathcal{N}_\mathsf{N}} \beta_i s_{i} + v_k,
\end{align}
where $v_k$ is zero-mean complex Gaussian additive white noise with variance $N_0$.

After receiving the signal in \eqref{eq:k-th_user_Rx_signal}, the $k$-th user first decodes messages of all weaker users (i.e., all $m$-th users such that $m \,{>}\, k$) sequentially in the presence of all the relatively stronger users' messages (i.e., those allocated less power). The decoded messages are then subtracted from the received signal of \eqref{eq:k-th_user_Rx_signal} in sequence, and the $k$-th user decodes its own message treating the stronger users' messages as noise (i.e., all $m$-th users' messages with $m \,{<}\, k$) . This overall decoding process is referred to as successive interference cancellation (SIC). Assuming that all the interfering messages of the users weaker than the $m$-th user are decoded and subtracted \textit{perfectly} at the $k$-th user, the signal-to-interference-plus-noise ratio (SINR) while decoding the $m$-th user message (at the $k$-th user) is given for $m \,{\geq}\, k$ as 
\begin{align}\label{eq:sinr_user}
    \mathsf{SINR}_{m{\rightarrow}k}^\mathsf{U} = \frac{\mathsf{P}_\mathsf{tx} |\textbf{h}_{k}^{\rm H}\textbf{w}|^2 \beta_{m}^2}{ \sum \limits_{l < m,\, l \in \mathcal{N}_\mathsf{N}} \mathsf{P}_\mathsf{tx} |\textbf{h}_{k}^{\rm H}\textbf{w}|^2 \beta_{l}^2 + N_0},
\end{align} 
which yields $\mathsf{SINR}_{k{\rightarrow}k}^\mathsf{U} \,{=}\, \mathsf{P}_\mathsf{tx} | \textbf{h}_{k}^{\rm H}\textbf{w}|^2 \beta_{k}^2 / N_0$ when $m \,{=}\, k$ represents the index of the strongest user. 

Note that the term $|\textbf{h}_{k}^{\rm H}\textbf{w}|^2$ is sufficient to characterize the SINR in \eqref{eq:sinr_user}, which involves the contribution of not only the individual user channels, but also the precoding vector, and is referred to as \textit{effective channel gain}. Assuming critically spaced ULA (i.e., $d_\mathsf{s} \,{=}\, \lambda_\mathsf{c}/2$), the effective channel gain for the $k$-th legitimate user is given using \eqref{eq:k_UE_original_channel} as follows~\cite{Ding17PoorRandBeamforming}
\begin{align} 
|\textbf{h}_k^{\rm H}\textbf{w}|^2 &\approx \frac{|\alpha_k|^2}{M \times\mathsf{PL}\left(\sqrt{d_k^2 + h^2}\right)}
\left| \frac{ \sin \left( \frac{\pi M(\overline{\theta} - \theta_k)}{2} \right)}{  \sin \left( \frac{\pi (\overline{\theta} - \theta_k)}{2} \right)}\right|^2 \nonumber \\
&= \frac{|\alpha_k|^2}{\mathsf{PL}\left(\sqrt{d_k^2 + h^2}\right)} \mathsf{F}_M \left(\overline{\theta}, \theta_k \right), \label{eq:effective_channel_gain}
\end{align} 
where $\mathsf{F}_M(\cdot)$ is called Fej\'er kernel function. 

\subsection{Decoding Performance at Eves} \label{subsec:decoding_performance_eve}

The decoding performance of the superposition-coded signal in \eqref{eq:superposition_code} at the most detrimental Eve (i.e., the one with the largest effective channel, or, equivalently, best channel condition towards the UAV-BS) depends on the presumed decoding capability of the Eves. In \cite{Ding2016SecSum, Yuan2019BeaDes, Lau2017OnDes, Qin2017SecSum}, it is assumed that the (most detrimental) Eve has already decoded the relatively weaker users' messages while decoding a specific user message by SIC. This approach implicitly assumes that the Eve has the knowledge of decoding order and power allocation for the legitimate users, which clearly overestimates the Eve's decoding capability. This approach is therefore referred to as the \textit{worst-case condition} (from the legitimate user's point of view), and is widely adopted in the literature to obtain a conservative \textit{lower bound} for the secrecy rates. In other words, realistic secrecy rate results can be \textit{reasonably} expected to be greater than this worst-case lower bound.   

It is, however, assumed in \cite{Ding2019SecTra, Ng2018ExpInt} that the most detrimental Eve does not apply SIC (because it may not have either decoding order or power allocation information for the legitimate NOMA users). The Eve rather treats all the other interfering legitimate NOMA user messages as noise (irrespective of belonging to relatively weaker or stronger users) while decoding any NOMA user message. This approach seemingly does not assume any multiuser decoding capability at the Eve, and is referred to as the \textit{best-case condition} (from the legitimate user's point of view), or \textit{upper bound} on the secrecy performance \cite{Tsiftsis2019SecUse}. 

\begin{remark}
Note that a second lower bound is also considered in \cite{Tsiftsis2019SecUse} for the secrecy-rate performance, which assumes that each legitimate user's message is detected at the Eve under the observation noise only (without any multiuser interference). This assumption leads only to a \textit{theoretical} performance bound for the secrecy rates (i.e., overly high decoding capability at the most detrimental Eve), which is even smaller than the lower bound associated with the worst-case condition (since the most detrimental Eve still takes into account the multiuser interference for the worst-case condition). Assuming that the decoding order and power allocation information is available to the most detrimental Eve, the respective decoding performance (at the most detrimental Eve) is somewhere between the worst- and best-case scenarios defined above. The actual decoding performance basically depends on the detection accuracy of the relatively weaker user's messages at the most detrimental Eve. Note also that as the most detrimental Eve decodes the weaker users' message erroneously, this will in turn appear as an additional interference for the subsequent detection iterations through \textit{error propagation}, which will eventually improve the overall secrecy rates at the legitimate users.    
\end{remark}

Assuming $g_\mathsf{E}$ as the effective channel gain of the most detrimental Eve, and replacing it with $|\textbf{h}_{k}^{\rm H}\textbf{w}|^2$ in \eqref{eq:sinr_user}, the SINR for the worst-case scenario while decoding the $m$-th user message at the most detrimental Eve is given by 
\begin{align}\label{eq:sinr_eve_worstcase}
    \mathsf{SINR}_{m}^\mathsf{E} = \frac{\mathsf{P}_\mathsf{tx} g_\mathsf{E} \beta_{m}^2}{ \sum \limits_{l < m,\, l \in \mathcal{N}_\mathsf{N}} \mathsf{P}_\mathsf{tx} g_\mathsf{E} \beta_{l}^2 + N_0},
\end{align} 
while that for the best-case scenario is  
\begin{align}\label{eq:sinr_eve_bestcase}
    \mathsf{SINR}_{m}^\mathsf{E} = \frac{\mathsf{P}_\mathsf{tx} \, g_\mathsf{E} \beta_{m}^2}{ \sum \limits_{l \neq m,\, l \in \mathcal{N}_\mathsf{N}} \mathsf{P}_\mathsf{tx} \, g_\mathsf{E} \beta_{l}^2 + N_0}.
\end{align} 

As for the channel description in \eqref{eq:k_UE_original_channel}, the effective channel gain of any Eve can be represented by \eqref{eq:effective_channel_gain} employing the appropriate channel parameters for the Eves (i.e., horizontal distance, small-scale fading, and AoD). We accordingly describe the effective channel gain of the most detrimental Eve as follows 
\begin{align} \label{eq:most_detrimental_eve}
g_\mathsf{E} = \max_{k \,{\in}\, \mathcal{N}_\mathsf{E}} |\textbf{g}_{k}^{\rm H}\textbf{w}|^2
\end{align} 
where $\textbf{g}_k$ is the channel gain of the $k$-th Eve. Note also that we assume that the UAV-BS has no channel state information (CSI) for the Eves while it has sufficient level of information on the channel qualities of the legitimate users (to be able to apply NOMA).

\subsection{Sum Secrecy Rates} \label{subsec:sum_secrecy_rates}

The SIC approach requires that message of any legitimate user should be decoded not only at its own user but also at every other legitimate user which has relatively better channel condition. Assuming that $\mathsf{R}_{k{\rightarrow}m}\,{=}\,\log(1\,+\,\mathsf{SINR}_{k{\rightarrow}m}^\mathsf{U})$ represents the instantaneous rate associated with the decoding of the $k$-th user's message at the $m$-th user, the achievable rate for the $k$-th user message is given as
\begin{align}\label{eq:rate_noma_1}
    \mathsf{R}_{k}^\mathsf{NOMA} = \min \Big\lbrace \mathsf{R}_{k{\rightarrow}m} \, \big| \, m \leq k, \; \forall \, m \in \mathcal{N}_\mathsf{N} \Big\rbrace.
\end{align}
Considering \eqref{eq:sinr_user}, we observe that $\mathsf{SINR}_{k{\rightarrow}m}^\mathsf{U} \,{\geq}\, \mathsf{SINR}_{k{\rightarrow}n}^\mathsf{U}$ for any $m,n \,{\in}\, \mathcal{N}_\mathsf{N}$ with $m \,{\leq}\, n$ since $|\textbf{h}_{m}^{\rm H}\textbf{w}|^2 \,{\geq}\, |\textbf{h}_{n}^{\rm H}\textbf{w}|^2$ by definition, and \eqref{eq:rate_noma_1} accordingly yields
\begin{align}\label{eq:rate_noma}
    \mathsf{R}_{k}^\mathsf{NOMA} = \log \left(1 + \mathsf{SINR}_{k{\rightarrow}k}^\mathsf{U} \right).
\end{align}

The achievable rate for decoding the $k$-th user message at the most detrimental Eve is 
\begin{align}\label{eq:rate_eve}
    \mathsf{R}_{k,\mathsf{E}}^\mathsf{NOMA} = \log \left(1 + \mathsf{SINR}_{k}^\mathsf{E} \right),
\end{align}
which applies to both the worst- and best-case conditions (regarding the Eve's decoding capability) with the adequate SINR expression of either \eqref{eq:sinr_eve_worstcase} or \eqref{eq:sinr_eve_bestcase}. Note that the assumption of having already decoded the weaker user's messages in the worst-case condition eliminates any necessity to consider other SINR terms (corresponding to the weaker users messages) in \eqref{eq:rate_eve}. The secrecy rate for the $k$-th legitimate user is therefore given as follows
\begin{align} \label{eq:rate_secrecy}
\mathsf{R}^\mathsf{NOMA}_{k,\mathsf{SEC}} = \Big[\mathsf{R}^\mathsf{NOMA}_k- \mathsf{R}^\mathsf{NOMA}_{k,\mathsf{E}}\Big]^+ \!\! ,
\end{align} 
where $[x]^{+}\,{=}\,\max\{x,0\}$. As \eqref{eq:rate_secrecy} implies, the secrecy rates are always nonnegative.   

Assuming that $\overline{\mathsf{R}}_k$ denotes the target secrecy rate for the $k$-th legitimate user, we define the respective conditional secrecy outage probability as follows
\begin{align}\label{eq:outage_secrecy}
\mathsf{P}_{k|K,K_\mathsf{e}}^\mathsf{o} = \mathsf{Pr} \left( \mathsf{R}^\mathsf{NOMA}_{k,\mathsf{SEC}} < \overline{\mathsf{R}}_k \mid K \in \mathcal{S}_\mathsf{N}, K_\mathsf{e} \geq 1 \right),
\end{align} 
where $\mathcal{S}_\mathsf{N}$ is the set of number of legitimate users for which NOMA is possible. Note that there should be at least $K_\mathsf{min}^\mathsf{N}$ legitimate users to make NOMA possible such that $K_\mathsf{min}^\mathsf{N} \,{\geq}\, \argmax_k \{k \,{|}\, k \,{\in}\, \mathcal{N}_\mathsf{N}\}$, and we therefore have $\mathcal{S}_\mathsf{N} \,{=}\, \{k \,{|}\, k \,{\geq}\, K_\mathsf{min}^\mathsf{N}, k \,{\in}\, \mathbb{Z}^+\}$. The outage-based sum secrecy rate for NOMA is then given as
\begin{align} \label{eq:rate_noma_secrecy}
\mathsf{R}^\mathsf{NOMA}_\mathsf{SEC} &= \sum\limits_{n \in \mathcal{S}_\mathsf{N}}^{\infty} \sum\limits_{m=1}^{\infty} \mathsf{Pr} \left( K=n \right)  \mathsf{Pr} \left( K_\mathsf{e}=m \right) \nonumber \\ 
& \hspace{0.8in} \times \Bigg( \sum\limits_{\substack{ k \in \mathcal{N}_\mathsf{N}}} \big( 1 - \mathsf{P}_{k|K,K_\mathsf{e}}^\mathsf{o} \big) \overline{\mathsf{R}}_k \Bigg),
\end{align} 
which assumes the presence of at least one Eve in the Eve region. When we have $K \,{<}\, K_\mathsf{min}^\mathsf{N}$, the NOMA transmission is not possible any more, and we resort to single user transmission (SUT) where the legitimate user with the best channel quality (among the ones in $\mathcal{N}_\mathsf{N}$) is scheduled during the whole course of transmission. Note that the number of legitimate users should satisfy $K_\mathsf{min}^\mathsf{S} \,{\leq}\, K \,{\leq}\, K_\mathsf{min}^\mathsf{N}$ for SUT such that $K_\mathsf{min}^\mathsf{S} \,{=}\, \argmin_k \{k \,{|}\, k \,{\in}\, \mathcal{N}_\mathsf{N}\}$ is the minimum number of legitimate users for SUT. Assuming $\mathcal{S}_\mathsf{S} \,{=}\, \{k \,{|}\, K_\mathsf{min}^\mathsf{S} \,{\leq}\, k \,{\leq}\, K_\mathsf{min}^\mathsf{N}, k \,{\in}\, \mathbb{Z}^+\}$, the respective SUT secrecy rate is given as  
\begin{align} \label{eq:rate_sut_secrecy}
\mathsf{R}^\mathsf{SUT}_\mathsf{SEC} &=  \sum\limits_{n \in \mathcal{S}_\mathsf{S}}^{\infty} \sum\limits_{m=1}^{\infty} \mathsf{Pr} \left( K=n \right)  \mathsf{Pr} \left( K_\mathsf{e}=m \right)  \nonumber \\
& \hspace{1.5in} \times \big( 1 - \mathsf{P}_{k|K,K_\mathsf{e}}^\mathsf{o} \big) \overline{\mathsf{R}}_k,
\end{align}
where $k$ is the index of the scheduled legitimate user, and
\begin{align}\label{eq:outage_secrecy_sut}
\mathsf{P}_{k|K,K_\mathsf{e}}^\mathsf{o} {=}\, \mathsf{Pr} \left( \left[\mathsf{R}^\mathsf{SUT}_k {-} \mathsf{R}^\mathsf{SUT}_{k,\mathsf{E}} \right]^{+} {<} \overline{\mathsf{R}}_k \,{\mid}\, K {\in}\, \mathcal{S}_\mathsf{S}, K_\mathsf{e} \geq 1 \right) \!, 
\end{align} 
with $\mathsf{R}^\mathsf{SUT}_k {=} \log \left( 1 {+} \frac{ \mathsf{P}_\mathsf{tx} }{ N_0 } | \textbf{h}_{k}^{\rm H} \textbf{w}|^2 \right)$ and $\mathsf{R}^\mathsf{SUT}_{k,\mathsf{E}} {=} \log \left( 1 {+} \frac{ \mathsf{P}_\mathsf{tx} }{ N_0 } \, g_\mathsf{E} \right)$. The final sum secrecy rate is given by \eqref{eq:rate_noma_secrecy} and \eqref{eq:rate_sut_secrecy} as
\begin{align}\label{eq:sum_secrecy_rate}
\mathsf{R}_\mathsf{SEC} \,{=}\, \mathsf{R}^\mathsf{NOMA}_\mathsf{SEC} \,{+}\, \mathsf{R}^\mathsf{SUT}_\mathsf{SEC}.
\end{align}

\subsection{Shape Optimization for Protected Zone} \label{subsec:shape_optimization}

In this section, we discuss optimization of the protected zone shape to enhance the secrecy rates while keeping its area the same. Considering the definition of the protected zone area in  \eqref{eq:area_pzone_I}-\eqref{eq:area_pzone_III}, it is obvious that a continuous set of $(\Delta_\mathsf{p},\, L_\mathsf{p})$ pairs produce the same area fraction $q$, all of which require the same ground resources. We therefore seek for the best angle-distance pair (or, equivalently, the best \textit{protected zone shape}) to maximize the secrecy rates. Note that each subregion of the Eve region with a particular angle-distance pair (i.e., candidate for the protected zone) does not impair the secrecy rates equally even if areas of these subregions are the same and the Eves are all equally capable. This is due to the fact that the effective channel gain between the UAV-BS and any Eve is a function of the Eve's location, and therefore takes different values in each subregion which represents a different channel quality. As a result, decoding success of any Eve changes based on the particular subregion that includes it.   

Thanks to the directional propagation feature of the mmWave links, the effective channel gain of the Eves is a function of not only the distance but also the \textit{relative angle} (i.e., angle offset from the beamforming direction) associated with each Eve. As a result, Eves with better effective channel gain, or equivalently being in specific \textit{unfavorable} subregions (i.e., represented by particular distance-angle pairs) impair the secrecy rates more. Indeed, the subregion involving the most detrimental Eve has the largest (negative) impact on the secrecy rates. Hence, instead of choosing the subregions arbitrarily to form the protected zone, it is more reasonable to include (i.e., \textit{protect}) subregions which yield better effective channel gain for Eves, and hence are likely to involve the most detrimental Eve.
 
We therefore consider to optimize the shape of the protected zone, or equivalently find the optimal $(\Delta_\mathsf{p},\, L_\mathsf{p})$ pair, to enhance secrecy rates while keeping its area the same. At a particular UAV-BS altitude and for a given area fraction $q$, the optimal shape can be formulated as follows 
\begin{IEEEeqnarray}{rl}
\max_{\Delta_\mathsf{p},\, L_\mathsf{p}} & \qquad \mathsf{R}_\mathsf{SEC} \label{eq:optim}\\
\text{s.t.} & \qquad \left(\Delta_\mathsf{p},L_\mathsf{p} \right) \text{ satisfies \eqref{eq:area_pzone_I}-\eqref{eq:area_pzone_III}}, \IEEEyessubnumber \\
& \qquad \mathsf{A}_\mathsf{P} \text{ is fixed}, \IEEEyessubnumber
\end{IEEEeqnarray}
where $\mathsf{R}_\mathsf{SEC}$ is the sum secrecy rate defined in  \eqref{eq:sum_secrecy_rate}.

\begin{remark} \label{remark:space_vs_shape_limited}
Note that the impact of the optimization of protected zone shape depends on the particular values of the area fraction $q \,{\in}\, (0,1)$ and expansion ratio $\kappa \,{>}\, 1$. For large $q$ or small $\kappa$ values, there is not enough degrees of freedom to alter the shape of the protected zone as the respective area can only be constructed for a limited range of $(\Delta_\mathsf{p},\, L_\mathsf{p})$ values. The corresponding secrecy-rate performance is mostly dominated by the available size of the Eve region into which the protected zone is inserted, and is therefore referred to as \textit{space-limited}. On the other hand, as $q$ becomes smaller or $\kappa$ becomes larger, we end up with an optimal angle-distance pair which maximizes the secrecy rates for a particular angle $\Delta_\mathsf{p}$ and distance $L_\mathsf{p}$ values that are not close to their respective limit values specified in \eqref{eq:area_pzone_I}-\eqref{eq:area_pzone_III} for different protected zone types. The respective secrecy-rate performance is therefore mostly dominated by the shape of the protected zone (i.e., the optimal $(\Delta_\mathsf{p},\, L_\mathsf{p})$ pair), and is hence referred to as \textit{shape-limited}.
\end{remark}

\section{Secrecy Sum Rates for Hybrid NOMA/SUT} 
\label{sec:secrecy_sum_rates}

In this section, we derive analytical expression for the secrecy sum rates of the hybrid NOMA/SUT transmission scheme described in Section~\ref{sec:noma_transmission}. Without any loss of generality, we consider two NOMA users though the results can be generalized to multiple NOMA users.

\subsection{Outage Probability} \label{subsec:preliminaries}

We assume that the NOMA transmission targets $i$-th and $j$-th users with $i\,{>}\,j$, which are therefore designated as the weaker and stronger users, respectively. In accordance with the discussion in Section~\ref{subsec:sum_secrecy_rates}, we have  $K_\mathsf{min}^\mathsf{N} \,{=}\, i$ and $K_\mathsf{min}^\mathsf{S} \,{=}\, j$ for this particular case, and the number of legitimate users $K$ should be in the sets $\mathcal{S}_\mathsf{N} \,{=}\, \{k \,{|}\, k \,{\geq}\, i, k \,{\in}\, \mathbb{Z}^+\}$ and  $\mathcal{S}_\mathsf{S} \,{=}\, \{k \,{|}\, j \,{\leq}\, k \,{<}\, i, k \,{\in}\, \mathbb{Z}^+\}$ for the NOMA and SUT schemes, respectively. In addition, the SUT scheme schedules the $j$-th user whenever $i$-th user is not present. 

The conditional secrecy outage probability in \eqref{eq:outage_secrecy} for NOMA can be elaborated as follows
\begin{align}
\mathsf{P}_{k|K,K_\mathsf{e}}^\mathsf{o} &= \mathsf{Pr} \left(  \left[\mathsf{R}^\mathsf{NOMA}_k - \mathsf{R}^\mathsf{NOMA}_{k,\mathsf{E}}\right]^{+} < \overline{\mathsf{R}}_k \mid \Phi \right), \label{eq:outage_noma_1} \\
&= \mathsf{Pr} \left( 0 < \overline{\mathsf{R}}_k, \mathsf{R}^\mathsf{NOMA}_k < \mathsf{R}^\mathsf{NOMA}_{k,\mathsf{E}}\mid \Phi \right) \nonumber \\
& \hspace{-0.5in} + \mathsf{Pr} \left\lbrace \mathsf{R}^\mathsf{NOMA}_k - \mathsf{R}^\mathsf{NOMA}_{k,\mathsf{E}} < \overline{\mathsf{R}}_k, \mathsf{R}^\mathsf{NOMA}_k > \mathsf{R}^\mathsf{NOMA}_{k,\mathsf{E}} \mid \Phi \right\rbrace , \label{eq:outage_noma_2}
\end{align}
where $\Phi$ stands for the given condition $\{K \in \mathcal{S}_\mathsf{N}, K_\mathsf{e} \geq 1\}$ in \eqref{eq:outage_secrecy}. After further simplification, and noting that $\overline{\mathsf{R}}_k$ takes always positive values, \eqref{eq:outage_noma_2} turns out to be
\begin{align} 
\mathsf{P}_{k\mid \Phi}^\mathsf{o} &= \mathsf{Pr} \left( \mathsf{R}^\mathsf{NOMA}_k < \mathsf{R}^\mathsf{NOMA}_{k,\mathsf{E}} \mid \Phi \right) \nonumber \\
& \quad + \mathsf{Pr} \left( \mathsf{R}^\mathsf{NOMA}_{k,\mathsf{E}} < \mathsf{R}^\mathsf{NOMA}_k < \mathsf{R}^\mathsf{NOMA}_{k,\mathsf{E}} + \overline{\mathsf{R}}_k \mid \Phi \right) , \label{eq:outage_noma_3} \\
&= \mathsf{Pr} \left( \mathsf{R}^\mathsf{NOMA}_k < \mathsf{R}^\mathsf{NOMA}_{k,\mathsf{E}} + \overline{\mathsf{R}}_k \mid \Phi \right) , \label{eq:outage_noma_4}
\end{align} 
which will be evaluated in the following to derive the analytical secrecy outage probabilities for the NOMA scheme considering the worst- and best-case conditions regarding the decoding capability of Eves. 

The conditional outage secrecy probability for the SUT scheme in \eqref{eq:outage_secrecy_sut} can be elaborated following a similar strategy, and is given as follows
\begin{align} 
\mathsf{P}_{k\mid \Phi}^\mathsf{o} &= \mathsf{Pr} \left( \mathsf{R}^\mathsf{SUT}_k < \mathsf{R}^\mathsf{SUT}_{k,\mathsf{E}} + \overline{\mathsf{R}}_k \mid \Phi \right) , \label{eq:outage_sut_1}
\end{align} 
where $\Phi$ stands for the respective given condition $\{K \in \mathcal{S}_\mathsf{S}, K_\mathsf{e} \geq 1\}$. Since we have single user in the SUT scheme scheduled all the time, there is no worst- or best-case conditions for decoding at the most detrimental Eve.

\begin{theorem}\label{theorem:outage_strong}
The conditional secrecy outage probability of the strong NOMA user in \eqref{eq:outage_noma_4}, and that of the SUT user in \eqref{eq:outage_sut_1} are given analytically as follows
\begin{align} \label{eq:outage_strong_sut}
\mathsf{P}_{j\mid \Phi}^\mathsf{o} = \int_{0}^{\infty} \int_{0}^{\delta^\mathsf{max}_j(y)} f_{\gamma_j \mid \Phi}(x) \, f_{g_\mathsf{E} \mid \Phi}(y) \, \dd x \dd y ,
\end{align}
where $\gamma_j \,{=}\, |\textbf{h}_{j}^{\rm H}\textbf{w}|^2$ is the effective channel gain for the $j$-th legitimate user, $f_{\gamma_j \mid \Phi}(x)$ and $f_{g_\mathsf{E} \mid \Phi}(y)$ are the PDF of the effective channel gain for the $j$-th legitimate user and most detrimental Eve, respectively, and the inner integration upper limit for the NOMA transmission is given as
\begin{subnumcases}{ \hspace{-0.2in} \delta^\mathsf{max}_j(y) \,{=} }
\frac{2^{\overline{\mathsf{R}}_j}{-}1}{ \frac{\mathsf{P}_\mathsf{tx}}{N_0} \beta_{j}^2}{+}2^{\overline{\mathsf{R}}_j}y, & \hspace{-0.2in} \text{worst-case}, \label{eq:upper_limit_worst}\\
\frac{2^{\overline{\mathsf{R}}_j}{-}1}{ \frac{\mathsf{P}_\mathsf{tx}}{N_0} \beta_{j}^2}+2^{\overline{\mathsf{R}}_j} \frac{y}{1 +\frac{\mathsf{P}_\mathsf{tx}}{N_0} y \beta_{i}^2}, & \hspace{-0.2in} \text{best-case}, \label{eq:upper_limit_best}
\end{subnumcases}
and $\delta^\mathsf{max}_j(y) \,{=}\, \frac{2^{\overline{\mathsf{R}}_j}{-}1}{ \frac{\mathsf{P}_\mathsf{tx}}{N_0} }{+}2^{\overline{\mathsf{R}}_j}y$ for SUT.
\end{theorem}
\begin{IEEEproof}
See Appendix~\ref{appendix:outage_strong}.
\end{IEEEproof}

\begin{remark} \label{remark:eve_decoding_capability}
As the transmit power becomes sufficiently large (i.e., $\mathsf{P}_\mathsf{tx} \,{\rightarrow}\, \infty$), the upper limit of the inner integral of \eqref{eq:outage_strong_sut} goes to zero (i.e., $\delta^\mathsf{max}_j \,{\rightarrow}\, 0$) for the best-case condition (where the most detrimental Eve has no multiuser decoding capability at all), and the outage probability in \eqref{eq:outage_strong_sut} accordingly becomes zero (i.e., $\mathsf{P}_{j\mid \Phi}^\mathsf{o} \,{\rightarrow}\, 0$). As a result, the strong NOMA user always achieves its target secrecy rate under large enough transmit power irrespective of the specific shape and size of the protected zone. However, this asymptotic behavior does not always hold for the worst-case scenario (which assumes that the Eve has already decoded the weak user message before decoding the strong user message) since $\delta^\mathsf{max}_j \,{\rightarrow}\, 2^{\overline{\mathsf{R}}_j} y$ as $\mathsf{P}_\mathsf{tx} \,{\rightarrow}\, \infty$ which may lead to a nonzero outage for the strong NOMA user. In Section~\ref{sec:results}, we show that the theoretical upper bound provided by the best-case scenario becomes very tight in high signal-to-noise ratio (SNR) regime (by approaching to the worst-case scenario) for sufficiently large protected zone (i.e., with large $q$) regardless of the particular power allocation (i.e., $\beta_j$ value) choice.
Note also that the integration upper limit $\delta^\mathsf{max}_j(y)$ for SUT can be obtained by assuming $\beta_i \,{=}\, 0$ and $\beta_j \,{=}\, 1$ in either \eqref{eq:upper_limit_worst} or \eqref{eq:upper_limit_best}, which aligns with the fact that the SUT scheme allocates all the resources to a single user (i.e., strong NOMA user) all the time.  
\end{remark}

\begin{theorem}\label{theorem:outage_weak}
The conditional secrecy outage probability of the weak NOMA user in \eqref{eq:outage_noma_4} is given analytically as follows
\begin{align} \label{eq:outage_weak}
\mathsf{P}_{i\mid \Phi}^\mathsf{o} &= \int_{0}^{\varrho} \int_{0}^{\delta_i(y)} f_{\gamma_{i} \mid \Phi}(x) \, f_{g_\mathsf{E} \mid \Phi}(y) \, \dd x \dd y \nonumber \\ 
& \qquad + \int_{\varrho}^{\infty} \int_{\delta_i(y)}^{\infty} f_{\gamma_{i} \mid \Phi}(x) \, f_{g_\mathsf{E} \mid \Phi}(y) \, \dd x \dd y ,
\end{align}
where $\gamma_i \,{=}\, |\textbf{h}_{i}^{\rm H}\textbf{w}|^2$ is the effective channel gain for the $i$-th legitimate user, $f_{\gamma_i \mid \Phi}(x)$ is the PDF of the effective channel gain for the $i$-th legitimate user, and the integration limits are
\begin{align}
\delta_i(y) &= \frac{2^{\overline{\mathsf{R}}_i}(1+\frac{\mathsf{P}_\mathsf{tx}}{N_0}y) - (1+\frac{\mathsf{P}_\mathsf{tx}}{N_0}\beta_{j}^2y) }{\frac{\mathsf{P}_\mathsf{tx}}{N_0} \left( (1+\frac{\mathsf{P}_\mathsf{tx}}{N_0}\beta_{j}^2y) - 2^{\overline{\mathsf{R}}_i}\beta_{j}^2(1+\frac{\mathsf{P}_\mathsf{tx}}{N_0}y)  \right)} \label{eq:threshold_weak} \\
\varrho &= \frac{ 1 - 2^{\overline{\mathsf{R}}_i} \beta_j^2 }{ \frac{\mathsf{P}_\mathsf{tx}}{N_0} \beta_j^2 \left( 2^{\overline{\mathsf{R}}_i} - 1 \right) } . \label{eq:outer_limit}
\end{align}

\end{theorem}
\begin{IEEEproof}
See Appendix~\ref{appendix:outage_weak}.
\end{IEEEproof}

\begin{remark} \label{remark:weak_noma}
Note that the integration limits given by \eqref{eq:threshold_weak} and \eqref{eq:outer_limit} both go to zero as the transmit power becomes \textit{sufficiently} large. In other words, $\delta_i(y) \,{\rightarrow}\, 0$ and $\varrho \,{\rightarrow}\, 0$ as $\mathsf{P}_\mathsf{tx} \,{\rightarrow}\, \infty$. Further elaborating \eqref{eq:outage_weak} by replacing both $\delta_i(y)$ and $\varrho$ with $0$, we observe that \eqref{eq:outage_weak} evaluates to $1$ for sufficiently large transmit power which implies that that the weak NOMA user is always in outage (in terms of secrecy rates) in this transmit power regime. In Section~\ref{sec:results}, we discuss the transmit power values after which the weak NOMA user is fully in outage through numerical examples.   
\end{remark}

\subsection{Distribution of Ordered Effective Channel Gain}

In order to compute the outage secrecy probabilities in \eqref{eq:outage_strong_sut} and \eqref{eq:outage_weak}, we need the \textit{ordered} PDF of the effective channel gain not only for the $i$-th and $j$-th legitimate users but also for the most detrimental Eve. By order statistics, the PDF of the effective channel gain for the $k$-th user with $k \,{\in}\, \{i,j\}$ can be expressed in terms of unordered distribution as follows  \cite{David03Order_stat} 
\begin{align} \label{eq:pdf_ordered_user}
\hspace{-0.0in} f_{|\textbf{h}_{k}^{\rm H}\textbf{w}|^2\mid \Phi}(z) = c_k \frac{{\rm d}F_{\mathsf{U}}(z)}{{\rm d}z}\left(F_{\mathsf{U}}(z)\right)^{K{-}k}\left(1{-}F_{\mathsf{U}}(z)\right)^{k{-}1},
\end{align} 
where $F_{\mathsf{U}}(z)$ is the CDF of unordered effective channel gain for any legitimate user with the given condition $\Phi$, and $c_k \,{=}\, \frac{K!}{(k{-}1)!(K{-}k)!}$. The PDF of the effective channel gain for the most detrimental Eve can also be given in a similar way as follows
\begin{align} \label{eq:pdf_ordered_eve}
f_{g_\mathsf{E}\mid \Phi}(y) = K_\mathsf{e} \frac{{\rm d}F_{\mathsf{E}}(y)}{{\rm d}y}\left(F_{\mathsf{E}}(y)\right)^{K_\mathsf{e}{-}1},
\end{align} 
where $F_{\mathsf{E}}(z)$ is the CDF of the unordered effective channel gain for any Eve. In the following, we provide the CDF of the unordered effective channel gain for the legitimate users, and derive that for the most detrimental Eve.

\begin{corollary}
The CDF of the unordered effective channel gain of any legitimate user for the scenario presented in Section~\ref{sec:system_model} is given as
\begin{align}\label{eq:cdf_unordered_user}
F_\mathsf{U}(z) \,{=} \int_{\overline{\theta}{-}\Delta_\mathsf{u}}^{\overline{\theta}{+}\Delta_\mathsf{u}} \hspace{-0.05in} \int_{L_\mathsf{i}}^{L_\mathsf{u}} \hspace{-0.05in} \Bigg( 1 \,{-}\, e^{ \frac{-z{\mathsf{PL} \big(\sqrt{r^2 + h^2}\big)}}{\mathsf{F}_M \left(\overline{\theta}, \theta_k \right) } }  \Bigg) \frac{r}{\mathsf{A}_\mathsf{u}} \dd r \dd \theta,
\end{align}
where $\mathsf{A}_\mathsf{u}\,{=}\, (L_\mathsf{u}^2 {-} L_\mathsf{i}^2) \Delta_\mathsf{u}$ is the area of the user region.
\end{corollary}
\begin{IEEEproof}
For the detailed proof, see \cite{Yapici2019AngFee}.
\end{IEEEproof}
Note that \eqref{eq:cdf_unordered_user} makes use of the fact that users are homogeneously distributed in the user region with the area $\mathsf{A}_\mathsf{u}$, and the respective PDF of the user location is therefore  $\frac{r}{\mathsf{A}_\mathsf{u}}$ in polar coordinates.

\begin{theorem}\label{theorem:cdf_unordered_eve}
The CDF of the unordered effective channel gain of any Eve for the scenario presented in Section~\ref{sec:system_model} is given as
\begin{align}\label{eq:cdf_unordered_eves}
\hspace{-0.1in} F_\mathsf{E}(y) \,{=}\, c \! \int_{\overline{\theta}{-}\Delta_\mathsf{e}}^{\overline{\theta}{+}\Delta_\mathsf{e}} \hspace{-0.05in} \int_{L_\mathsf{i}}^{L_\mathsf{e}} \hspace{-0.05in} \Bigg( \! 1 \,{-}\, e^{ \frac{-y{\mathsf{PL} \big(\sqrt{r^2 + h^2}\big)}}{ {\mathsf{F}_M \left(\overline{\theta}, \theta_k \right) } } }  \! \Bigg) I(\theta,r) r \dd r \dd \theta,
\end{align}
where $c \,{=}\, 1/(\mathsf{A}_\mathsf{e}{-}\mathsf{A}_\mathsf{p})$ with $\mathsf{A}_\mathsf{e} \,{=}\, \left( L_\mathsf{e}^2 \,{-}\, L_\mathsf{i}^2\right) \Delta_\mathsf{e} \,{-}\, \mathsf{A}_\mathsf{u}$ being the area of the Eve region and $\mathsf{A}_\mathsf{p}$ being the area of the protected zone given in \eqref{eq:area_pzone_I}-\eqref{eq:area_pzone_III}, and $I(\theta,r)$ is the indicator function given as
\begin{align}\label{eq:indicator}
    I(r,\theta) = \begin{cases}
    1 , & \text{ if } (\theta,r) \in \mathcal{S}_\mathsf{EP} ,\\
    0 , & \text{ otherwise},
    \end{cases}
\end{align}
with $\mathcal{S}_\mathsf{EP}$ being the set of all angle-distance pairs within the Eve region after the protected zone of interest is subtracted, which we refer to as \textit{unprotected-Eve region}.
\end{theorem}
\begin{IEEEproof}
See Appendix~\ref{appendix:cdf_unordered_eve}.
\end{IEEEproof}
 
\section{Numerical Results} \label{sec:results}

In this section, we present numerical results based on extensive Monte Carlo simulations to evaluate the secrecy performance of the hybrid NOMA/SUT transmission strategy under consideration. We numerically verify the analytical expressions derived in Section~\ref{sec:secrecy_sum_rates}, as well, which are shown to have a very good match with the simulation data. We consider the frequency-dependent mmWave PL model adopted by 3GPP urban micro (UMi) environment given as \cite{3GPP_TR38901} $$\mathsf{PL}(x) \,{=}\, 32.4 \,{+}\, 21\log_{10}(x) \,{+}\, 20\log_{10}(f_{\rm c}),$$ where $x$ is the LoS distance, and $f_{\rm c}$ is the carrier frequency (normalized by $1\,\text{GHz}$). The noise power is given as $N_0 \,{=}\, \mathsf{TNP} \,{+}\, 10 \log_{10}(\mathsf{BW}) \,{+}\, \mathsf{NF}$, where $\mathsf{TNP}$ is the thermal noise power, $\mathsf{BW}$ is the transmission bandwidth, and $\mathsf{NF}$ is the noise figure at the receive end. In order to have sufficiently distinct strong and weak users in the power domain, we choose the NOMA user indices as $i \,{=}\, 10$ and $j \,{=}\, 1$. The complete list of simulation parameters are given in Table~\ref{tab:simulation_parameters}.

\begin{table}[!t]
    \renewcommand{\arraystretch}{1.2}
	\caption{Simulation Parameters}
	\label{tab:simulation_parameters}
	\centering
	\begin{tabular}{lc}
	\hline
	Parameter & Value \\
	\hline\hline
    User-region range, $[L_\mathsf{i},L_\mathsf{u}]$ & $[5,50]\,\text{m}$ \\
    User-region (half) angle, $\Delta_\mathsf{u}$  & $2.5^{\circ}$ \\
    User density, $\lambda_\mathsf{u}$ & $1$ \\
    Eve density, $\lambda_\mathsf{e}$ & $0.1$ \\
    Number of BS antennas, $M$ & $100$ \\
    Maximum antenna gain at BS & $8\,\text{dBi}$ \\
    Thermal noise power, $\mathsf{TNP}$ & $-174\,\text{dBm/Hz}$ \\
    Transmission bandwidth, $\mathsf{BW}$ & $100\,\text{MHz}$ \\
    Noise figure, $\mathsf{NF}$ & $9\,\text{dB}$ \\
    Frequency, $f_\mathsf{c}$ & $28\,\text{GHz}$ \\
    Target secrecy rates, $\{\overline{\mathsf{R}}_i,\overline{\mathsf{R}}_j\}$ & $\{1,4\}\,\text{BPCU}$ \\
    Power allocation ratios, $\{\beta_i^2,\beta_j^2\}$ & $\{0.75,0.25\}$ \\
    UAV-BS operation altitude, $h$ & $\{20,50\}\,\text{m}$ \\
    \hline
	\end{tabular}
\end{table}

\begin{figure}[!t]
\centering
\subfloat[Sum secrecy rate vs. protected zone angle] {\includegraphics[width=0.51\textwidth]{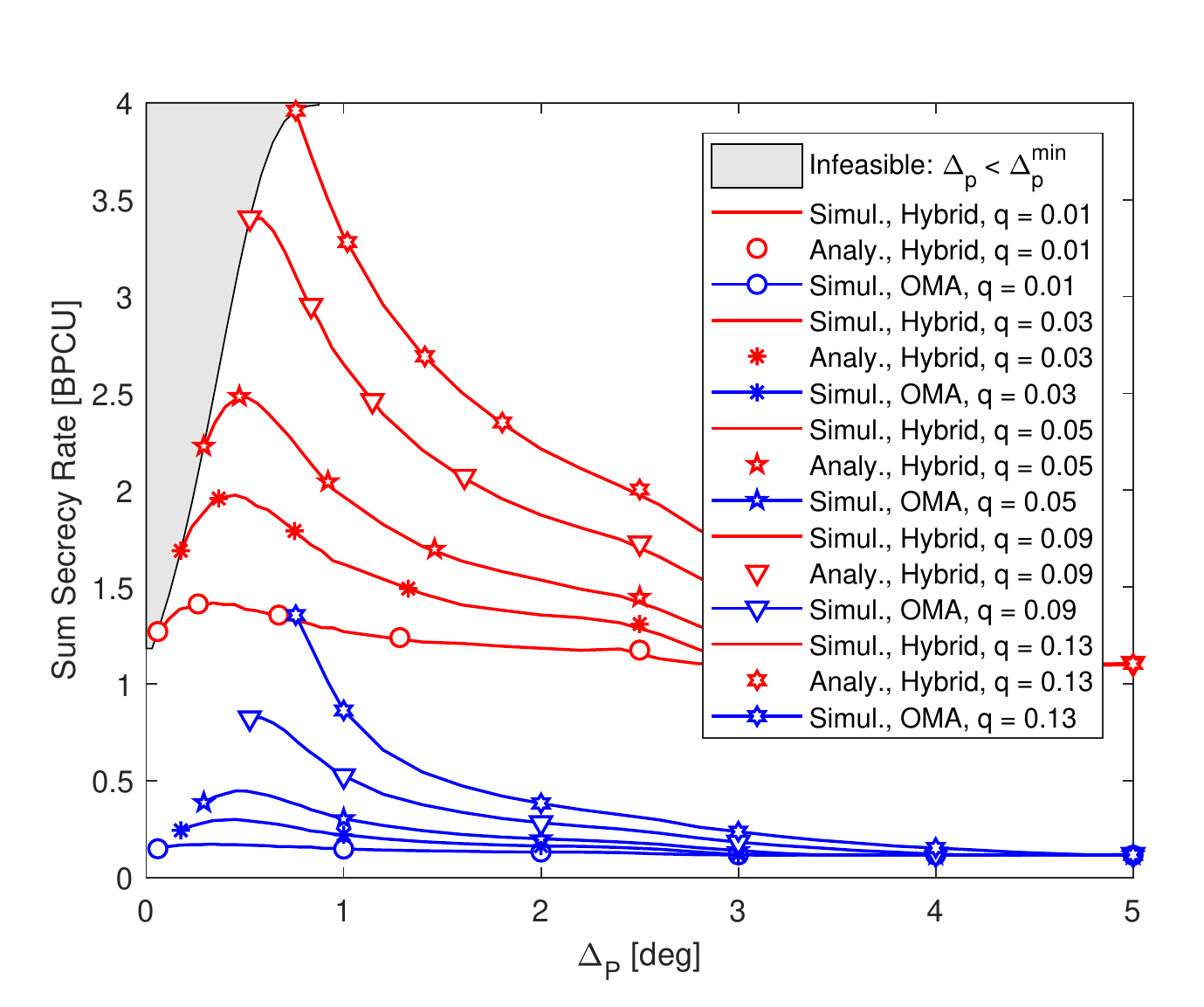}
\label{fig:rate_vs_angle_del5_h20m_rho2}}\\
\vspace{0.0in}
\subfloat[Sum secrecy rate vs. protected zone radius] {\includegraphics[width=0.51\textwidth]{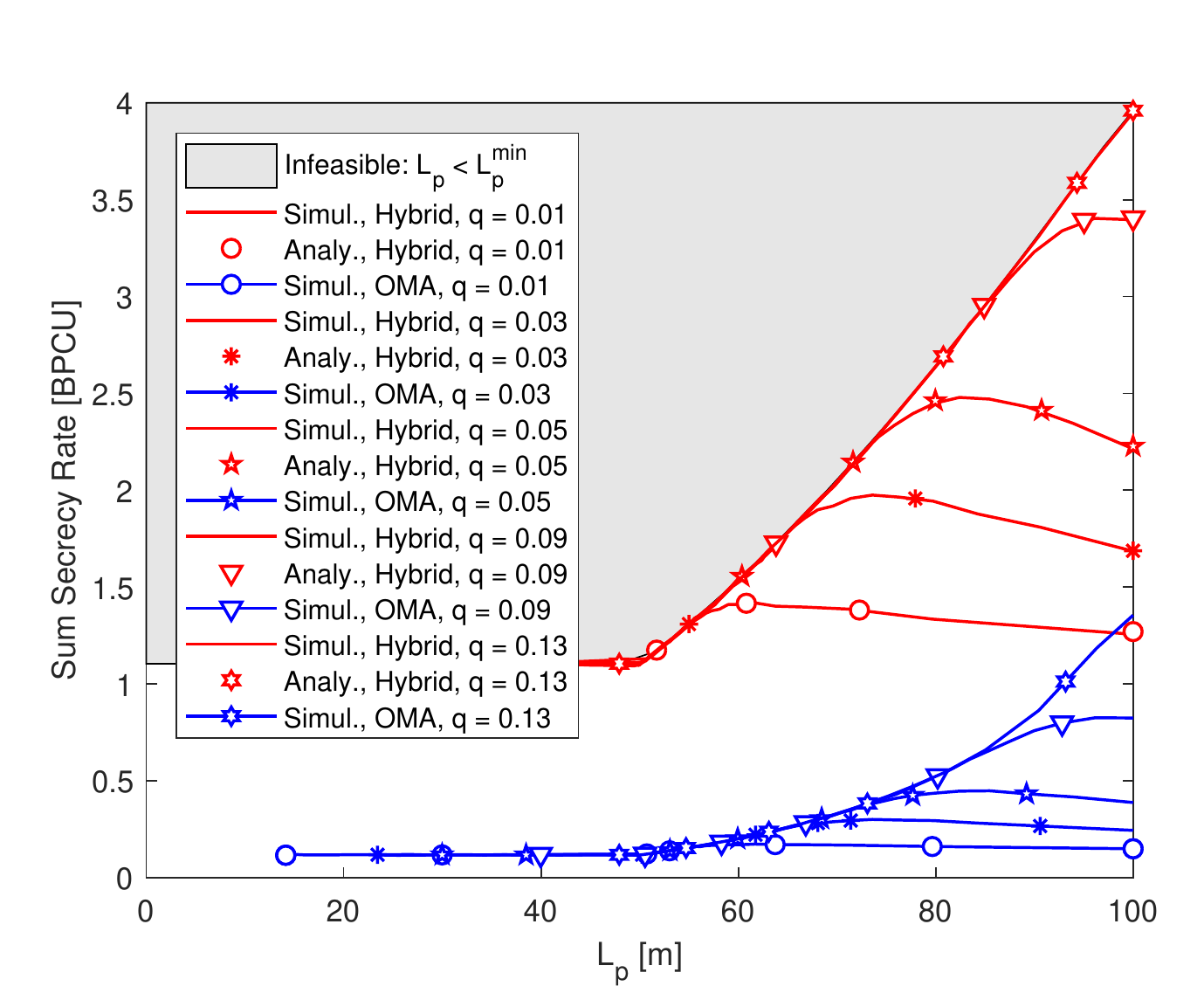}
\label{fig:rate_vs_distance_del5_h20m_rho2}}
\caption{Sum secrecy rate vs. protected zone shape (i.e., angle, radius) assuming worst-case Eve decoding, area fraction of $q \,{\in}\, [0.01,0.13]$, altitude of $h\,{=}\,20\,\text{m}$, expansion ratio of $\kappa\,{=}\,2$ (i.e., $\Delta_\mathsf{e}\,{=}\,2 \Delta_\mathsf{u}$, $L_\mathsf{e}\,{=}\,2 L_\mathsf{u}$), and transmit power of $\mathsf{P}_\mathsf{tx}\,{=}\,45 \,\text{dBm}$.}
\label{fig:rate_del5_h20m_rho2}
\vspace{-0.0in}
\end{figure}

In Fig.~\ref{fig:rate_del5_h20m_rho2}, we depict the sum secrecy rate along with varying protected zone shape (through $(\Delta_\mathsf{p},L_\mathsf{p})$ pairs) for the area fraction of $q \,{\in}\, [0.01,0.13]$. In particular, we assume the expansion ratio of $\kappa\,{=}\,2$ (i.e., $\Delta_\mathsf{e}\,{=}\,2 \Delta_\mathsf{u}$, $L_\mathsf{e}\,{=}\,2 L_\mathsf{u}$), UAV-BS altitude of $h\,{=}\,20\,\text{m}$, transmit power of $\mathsf{P}_\mathsf{tx}\,{=}\,45 \,\text{dBm}$, and worst-case Eve decoding. We observe that the secrecy performance of the hybrid NOMA/SUT scheme is superior to that of OMA, and is generally better for protected zone shapes having relatively small $\Delta_\mathsf{p}$ and large $L_\mathsf{p}$. This result implies that Fej\'er kernel in \eqref{eq:effective_channel_gain}, which represents the impact of the angular offset on the effective channel gain of the most detrimental Eve, is dominant over the path loss. The optimal protected zone, therefore, takes a shape so as to cover the subregions (of the Eve region) with small angle-offset values (i.e., Type-I in Fig.~\ref{fig:protected_zone}), for which the effective channel gain of the Eves would be large. The sub-6GHz propagation, on the contrary, would require the subregions with small distance values to be protected (e.g., Type-II) from the presence of Eves since the underlying channel is rich-scattering, and path loss is hence the only feature to contribute to the average channel quality (of the most detrimental Eve).

We also observe in Fig.~\ref{fig:rate_del5_h20m_rho2} that the sum secrecy rate of the hybrid NOMA/SUT scheme is maximized for particular choices of $(\Delta_\mathsf{p},L_\mathsf{p})$ pairs, or, equivalently, protected zone shapes, for each $q$ value (i.e., fixed protected zone area). Note that the optimal $\Delta_\mathsf{p}$ (i.e., $\Delta_\mathsf{p}^\mathsf{opt}$) is greater than its minimum possible value $\Delta_\mathsf{p}^\mathsf{min}$ for $q \,{\in}\, \{0.01,0.03,0.05\}$, thereby representing a compromise between \textit{protecting} angle- versus distance-dimension (i.e., small $\Delta_\mathsf{p}$ vs. small $L_\mathsf{p}$). For this range of $q$ values, we have enough degrees of freedom in choosing $(\Delta_\mathsf{p},L_\mathsf{p})$ pairs while maximizing the secrecy sum rate (i.e., seeking for the optimal protected zone shape). The respective secrecy performance is therefore referred to as \textit{shape-limited} such that $\Delta_\mathsf{p}^\mathsf{min} \,{<}\, \Delta_\mathsf{p}^\mathsf{opt}\,{<}\, \Delta_\mathsf{e}$, as discussed in Remark~\ref{remark:space_vs_shape_limited}. 

On the other hand, as we desire to protect larger portion of the Eve region through greater area fraction of $q \,{\in}\, \{0.09,0.13\}$, $\Delta_\mathsf{p}^\mathsf{opt}$ becomes equal to $\Delta_\mathsf{p}^\mathsf{min}$. In this situation, the requested relatively-larger protected zone can only be constructed using a relatively limited range of $(\Delta_\mathsf{p},L_\mathsf{p})$ values, and corresponding $\Delta_\mathsf{p}^\mathsf{min}$ is larger accordingly as compared to those associated with small $q$ values. In this case, the optimal angular width of the protected zone is limited by $\Delta_\mathsf{p}^\mathsf{min}$ (i.e., $\Delta_\mathsf{p}^\mathsf{opt} \,{=}\, \Delta_\mathsf{p}^\mathsf{min}$). The respective secrecy performance is therefore referred to as \textit{space-limited} in the sense that the Eve region is not sufficiently large any more (as compared to the protected zone), which therefore loses the flexibility in choosing the optimal $(\Delta_\mathsf{p},L_\mathsf{p})$ pair to maximize secrecy rates.

\begin{figure}[!t]
\centering
\includegraphics[width=0.51\textwidth]{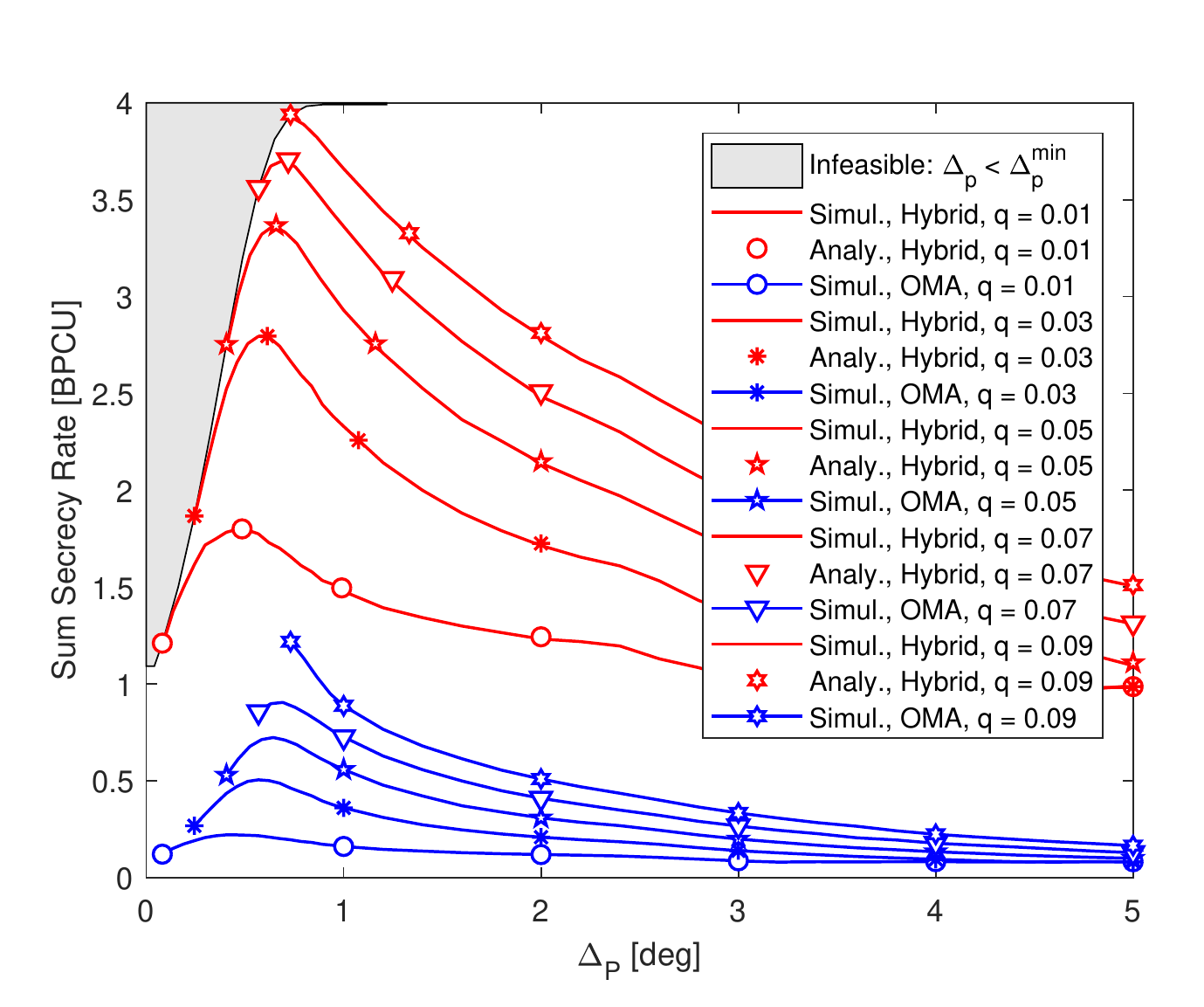}
\vspace{0.0in}
\caption{Sum secrecy rate vs. protected zone angle assuming worst-case Eve decoding, area fraction of $q \,{\in}\, [0.01,0.09]$, altitude of $h\,{=}\,20\,\text{m}$, expansion ratio of $\kappa\,{=}\,3$ (i.e., $\Delta_\mathsf{e}\,{=}\,3 \Delta_\mathsf{u}$, $L_\mathsf{e}\,{=}\,3 L_\mathsf{u}$), and transmit power of $\mathsf{P}_\mathsf{tx}\,{=}\,45 \,\text{dBm}$.}
\label{fig:rate_vs_angle_del5_h20m_rho3}
\vspace{-0.0in}
\end{figure}

In Fig.~\ref{fig:rate_vs_angle_del5_h20m_rho3}, we provide the sum secrecy rate results for a larger Eve region with the expansion ratio of $\kappa \,{=}\, 3$ keeping all the other simulation parameters of Fig.~\ref{fig:rate_del5_h20m_rho2} the same. We observe that the respective characteristic of the secrecy-rate performance is similar to that of Fig.~\ref{fig:rate_del5_h20m_rho2}. In order to better investigate the impact of larger $\kappa$ on the optimization of the protected zone shape, we focus on the \textit{space-limited} secrecy performance associated with $q \,{=}\, 0.09$ in Fig.~\ref{fig:rate_del5_h20m_rho2}, for which the best sum secrecy rate for the hybrid NOMA/SUT scheme is achieved when $\Delta_\mathsf{p}^\mathsf{opt} \,{=}\, \Delta_\mathsf{p}^\mathsf{min}$. Note that we need to consider $q \,{\approx}\, 0.025$ for $\kappa \,{=}\, 3$ to have the same protected zone area (in size), which can be reasonably well approximated by $q \,{=}\, 0.03$ in Fig.~\ref{fig:rate_vs_angle_del5_h20m_rho3}. We observe that the optimal secrecy performance of $q \,{=}\, 0.03$ ($\kappa \,{=}\, 3$) for the hybrid NOMA/SUT scheme is obtained when $\Delta_\mathsf{p}^\mathsf{opt} \,{>}\, \Delta_\mathsf{p}^\mathsf{min}$, and the respective secrecy performance turns out to be \textit{shape-limited}, as discussed in Remark~\ref{remark:space_vs_shape_limited}. More specifically, choosing $\Delta_\mathsf{p}$ optimally ends up with a sum secrecy rate which is roughly $22\%$ better than that for $\Delta_\mathsf{p}^\mathsf{min}$ in this setting. We therefore conclude that although $\Delta_\mathsf{p}^\mathsf{min}$ might be considered as a heuristic choice for $\Delta_\mathsf{p}^\mathsf{opt}$ in mmWave communications (aiming at protecting mostly the angle dimension), larger Eve regions (i.e., bigger $\kappa$) provide better flexibility in optimizing the shape of the protected zone which ends up with $\Delta_\mathsf{p}^\mathsf{opt} \,{>}\, \Delta_\mathsf{p}^\mathsf{min}$.

\begin{figure}[!t]
\centering
\includegraphics[width=0.51\textwidth]{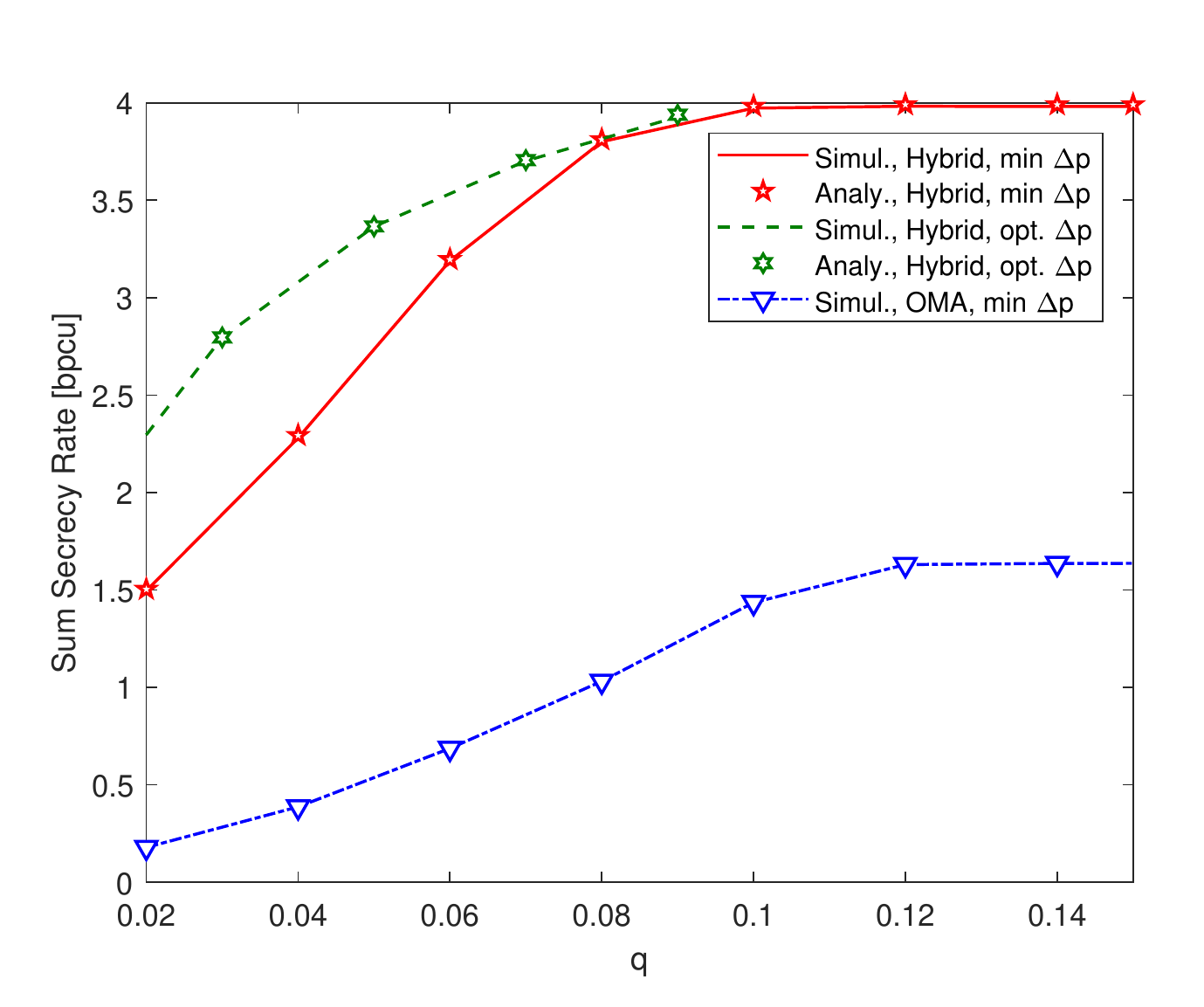}
\vspace{0.0in}
\caption{Sum secrecy rate vs. area fraction assuming worst-case Eve decoding, altitude of $h\,{=}\,20\,\text{m}$, expansion ratio of $\kappa\,{=}\,3$ (i.e., $\Delta_\mathsf{e}\,{=}\,3 \Delta_\mathsf{u}$, $L_\mathsf{e}\,{=}\,3 L_\mathsf{u}$), and transmit power of $\mathsf{P}_\mathsf{tx}\,{=}\,45 \,\text{dBm}$.}
\label{fig:rate_vs_Q_del5_h20m_rho3}
\vspace{-0.0in}
\end{figure}

We investigate the secrecy-rate convergence for varying area fraction $q$ in Fig.~\ref{fig:rate_vs_Q_del5_h20m_rho3} keeping all the other simulation parameters of Fig.~\ref{fig:rate_vs_angle_del5_h20m_rho3} the same. We observe that the difference of sum secrecy rates between $\Delta_\mathsf{p}^\mathsf{opt} \,{=}\, \Delta_\mathsf{p}^\mathsf{min}$ and $\Delta_\mathsf{p}^\mathsf{opt} \,{>}\, \Delta_\mathsf{p}^\mathsf{min}$ diminishes for the hybrid NOMA/SUT scheme as $q$ increases, thereby shifting the secrecy-rate performance from shape-limited to space-limited. In other words, seeking for an optimal $\Delta_\mathsf{p}$ loses its importance along with larger protected zone areas since $\Delta_\mathsf{p}^\mathsf{opt} \,{=}\, \Delta_\mathsf{p}^\mathsf{min}$ turns out to be already fair enough (since the protected zone cannot take any shape due to the available space in the Eve region being limited). We furthermore observe that the sum secrecy of the hybrid scheme rate saturates around $4$ bits per channel use (BPCU) after $q \,{=}\, 0.1$, which implies that \textit{allocating additional resources on the ground (i.e., choosing a larger protected zone) to protect more area within the Eve region is not necessary once the protected zone has a sufficiently reasonable size}. 

\begin{figure}[!t]
\centering
\subfloat[Sum secrecy rate] {\includegraphics[width=0.51\textwidth]{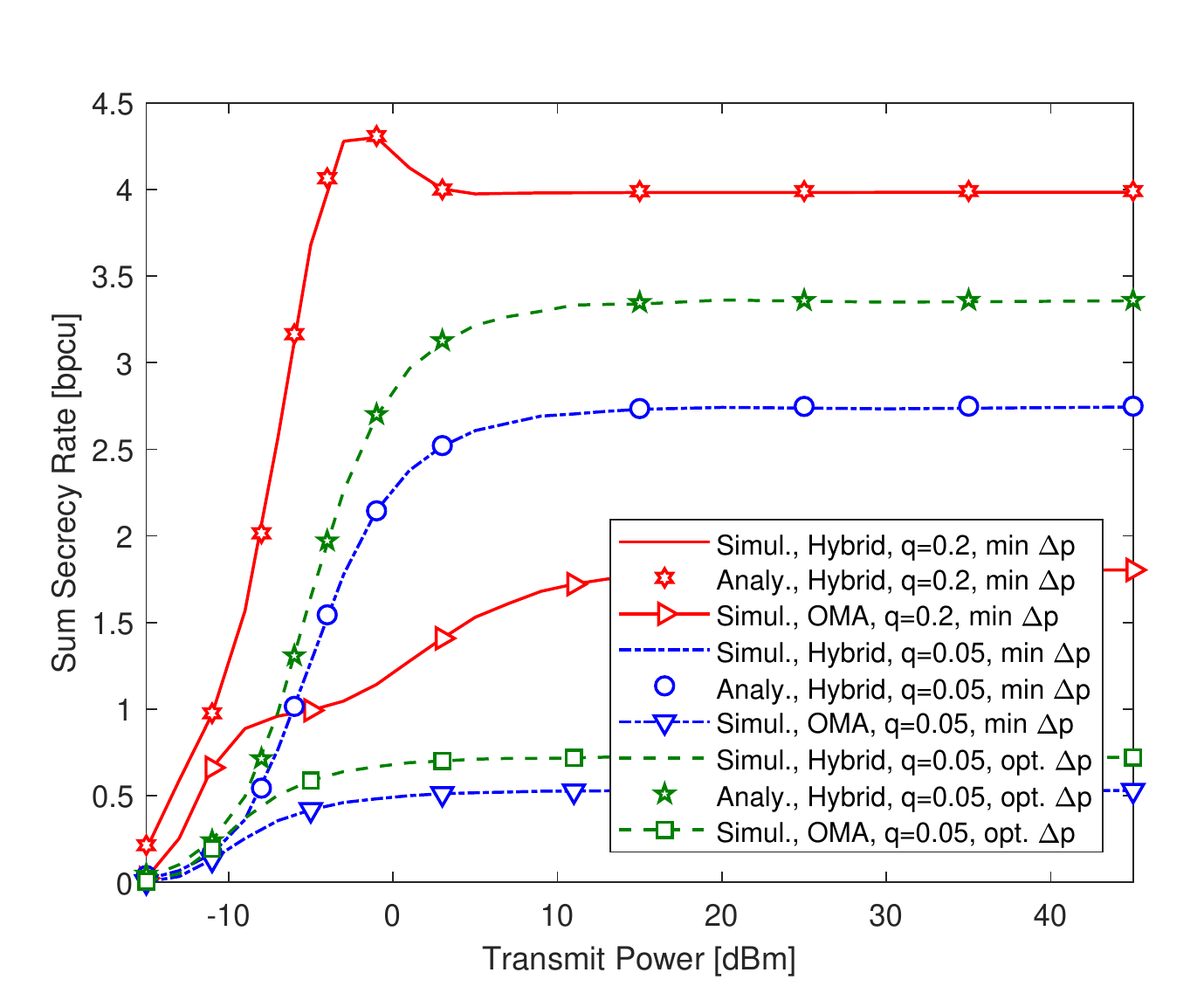}
\label{fig:rate_vs_Ptx_del5_h20m_rho3}}\\
\vspace{0.0in}
\subfloat[Outage probability] {\includegraphics[width=0.51\textwidth]{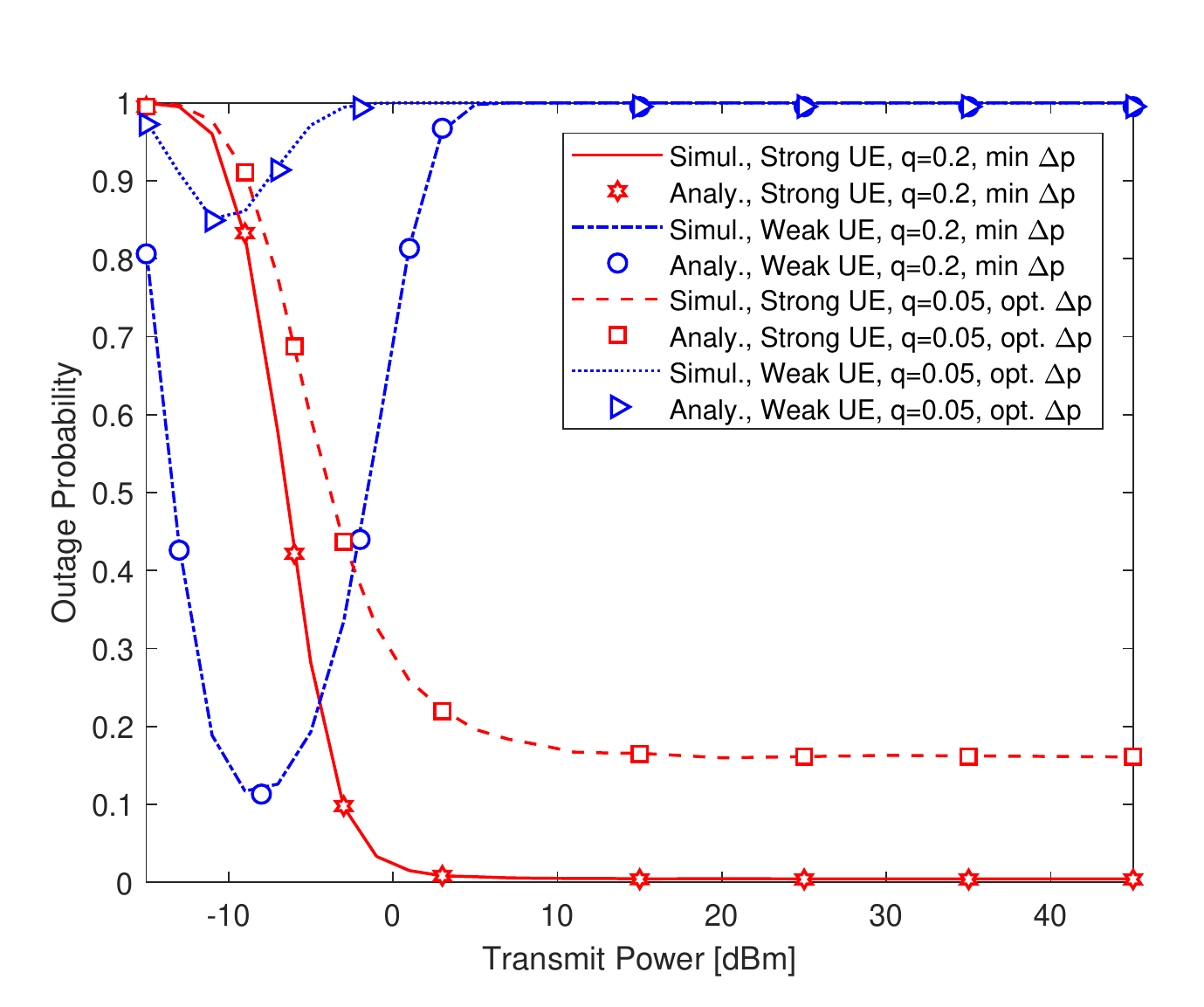}
\label{fig:outage_vs_Ptx_del5_h20m_rho3}}
\caption{Sum secrecy rate vs. transmit power assuming worst-case Eve decoding, area fraction of $q \,{\in}\, \{0.05,0.2\}$, altitude of $h\,{=}\,20\,\text{m}$, expansion ratio of $\kappa\,{=}\,3$ (i.e., $\Delta_\mathsf{e}\,{=}\,3 \Delta_\mathsf{u}$, $L_\mathsf{e}\,{=}\,3 L_\mathsf{u}$).}
\label{fig:Ptx_del5_h20m_rho3}
\vspace{-0.0in}
\end{figure}

In Fig.~\ref{fig:Ptx_del5_h20m_rho3}, we depict the secrecy-rate performance for varying transmit power of $\mathsf{P}_\mathsf{tx} \,{\in}\, [{-}15,45]$ dBm while keeping all the other simulation parameters of Fig.~\ref{fig:rate_vs_angle_del5_h20m_rho3} the same. In particular, we consider area fraction values of $q \,{=}\, 0.05$ and $q \,{=}\, 0.2$ which correspond to shape- and space-limited secrecy performance, respectively, as can be identified in Fig.~\ref{fig:rate_vs_Q_del5_h20m_rho3}. We observe in Fig.~\ref{fig:Ptx_del5_h20m_rho3}\subref{fig:rate_vs_Ptx_del5_h20m_rho3} that the sum secrecy rates saturate roughly after $\mathsf{P}_\mathsf{tx} \,{=}\, 5$ dBm for the hybrid NOMA/SUT scheme with either $q$ value and $\Delta_\mathsf{p} \,{\in}\, \left\lbrace \Delta_\mathsf{p}^\mathsf{opt},  \Delta_\mathsf{p}^\mathsf{min} \right\rbrace$. As depicted in Fig.~\ref{fig:Ptx_del5_h20m_rho3}\subref{fig:outage_vs_Ptx_del5_h20m_rho3}, the weak user is always in outage for the transmit power larger than  $\mathsf{P}_\mathsf{tx} \,{=}\, 5$ dBm, which is the reason for the maximum value of the sum secrecy rate after saturation being at most the target secrecy rate of the strong user (i.e., $\overline{\mathsf{R}}_j \,{=}\, 4$ BPCU). Note that the SINR expressions in \eqref{eq:sinr_user} and \eqref{eq:sinr_eve_worstcase} (associated with decoding the weak user message) converge to the same value (i.e., $\beta_i^2{/}\beta_j^2$) at high transmit power regime (i.e., $N_0/\mathsf{P}_\mathsf{tx} \,{\rightarrow}\, 0$). As a result, weak user cannot achieve any positive secrecy rate at high $\mathsf{P}_\mathsf{tx}$, and is always in outage since the respective nonzero target secrecy rate cannot be met. 

On the other hand, Fig.~\ref{fig:Ptx_del5_h20m_rho3}\subref{fig:rate_vs_Ptx_del5_h20m_rho3} depicts also that the sum secrecy rate makes a peak at $\mathsf{P}_\mathsf{tx} \,{=}\, {-}1.5$ dBm for $q \,{=}\, 0.2$, thereby exceeding the strong user target secrecy rate of $\overline{\mathsf{R}}_j \,{=}\, 4$ BPCU. Note that the SINR expressions in \eqref{eq:sinr_user} and \eqref{eq:sinr_eve_worstcase} do not converge to the same ratio of the power allocation coefficients in this transmit power regime, but rather remains as a function of the effective channel gains. Since the weak user has a sufficiently better channel quality than the most detrimental Eve for $q \,{=}\, 0.2$, its target secrecy rate is likely to be met with a nonzero probability, which results in an increase (i.e., peak) in the sum secrecy rates. Note that the effective channel gain of the weak user is not sufficiently better than that of the most detrimental Eve for $q \,{=}\, 0.05$ since the subregions that are likely to have the most detrimental Eve is not covered enough by the protected zone (due to small $q$). The weak user therefore cannot have as satisfactory outage performance as for $q \,{=}\, 0.2$, as depicted in Fig.~\ref{fig:Ptx_del5_h20m_rho3}\subref{fig:outage_vs_Ptx_del5_h20m_rho3}. As a final remark, we note that the sum secrecy rate of $q \,{=}\, 0.05$ with $\Delta_\mathsf{p}^\mathsf{opt}$ is superior to that with $\Delta_\mathsf{p}^\mathsf{min}$ for all transmit power levels.

\begin{figure}[!t]
\centering
\includegraphics[width=0.51\textwidth]{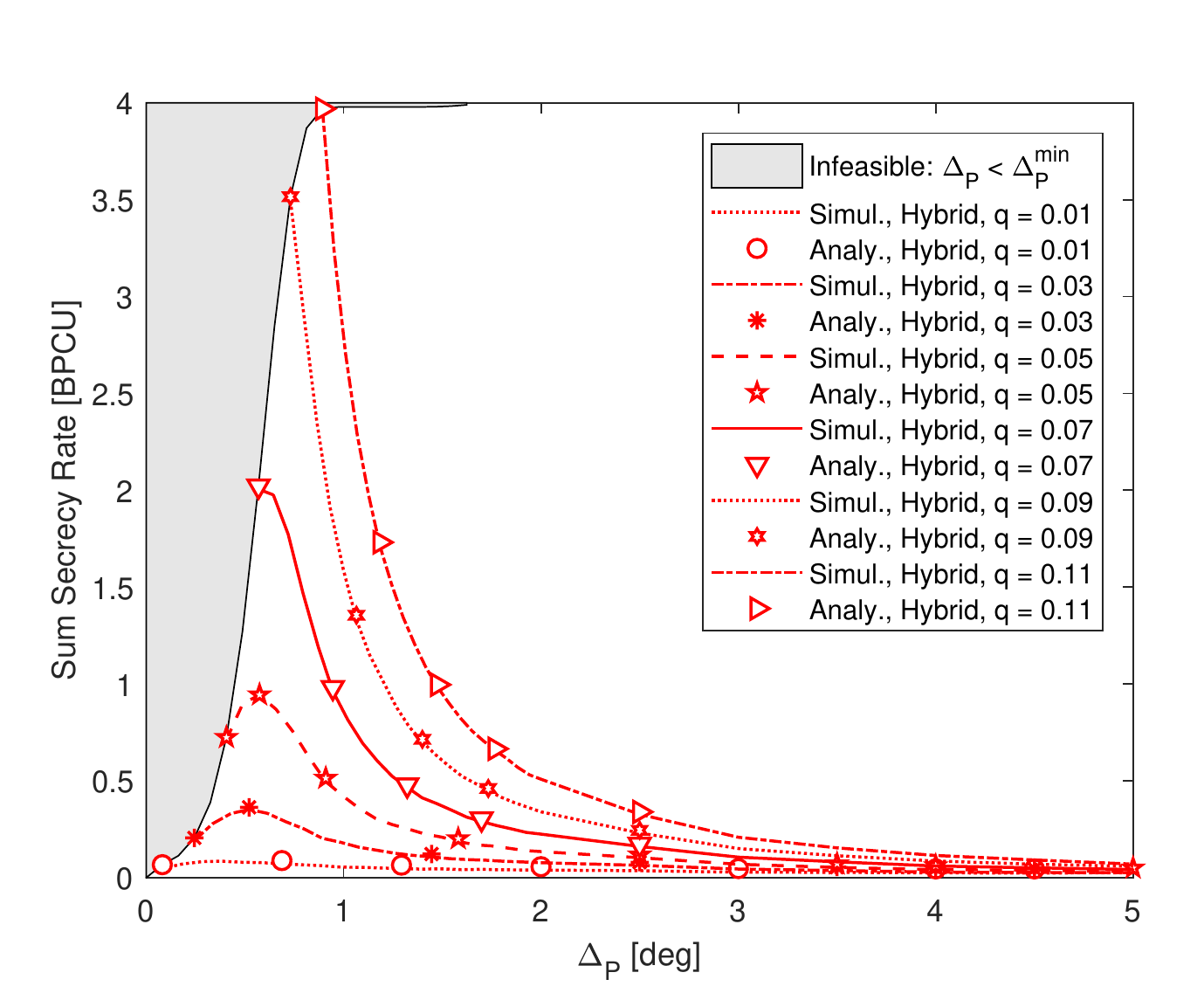}
\vspace{0.0in}
\caption{Sum secrecy rate vs. protected zone angle assuming worst-case Eve decoding, area fraction of $q \,{\in}\, [0.01,0.11]$, altitude of $h\,{=}\,50\,\text{m}$, expansion ratio of $\kappa\,{=}\,3$ (i.e., $\Delta_\mathsf{e}\,{=}\,3 \Delta_\mathsf{u}$, $L_\mathsf{e}\,{=}\,3 L_\mathsf{u}$), and transmit power of $\mathsf{P}_\mathsf{tx}\,{=}\,45 \,\text{dBm}$.}
\label{fig:rate_vs_angle_del5_h50m_rho3}
\vspace{0.0in}
\end{figure}

\begin{figure}[!t]
\centering
\includegraphics[width=0.51\textwidth]{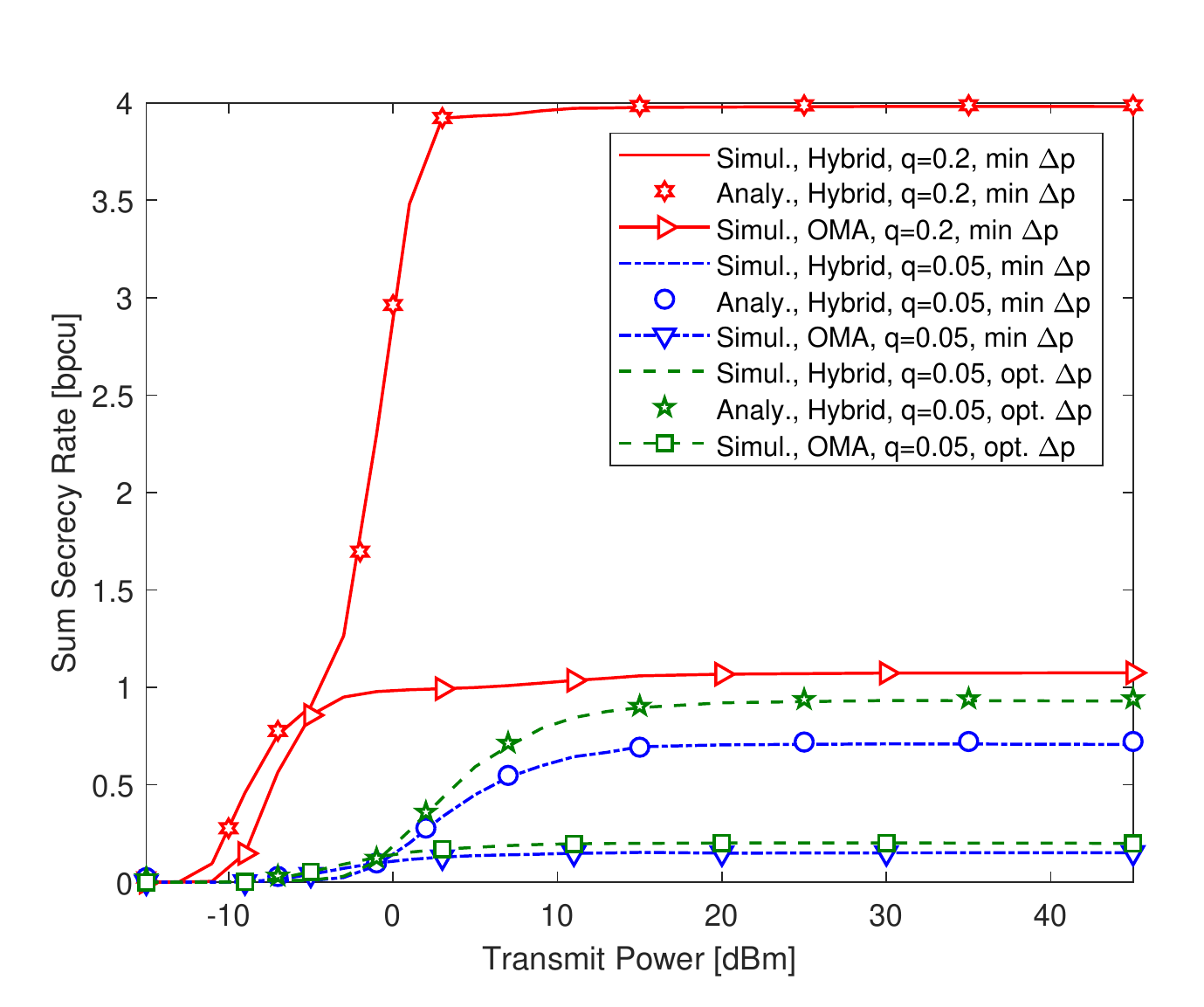}
\vspace{0.0in}
\caption{Sum secrecy rate vs. transmit power assuming worst-case Eve decoding, area fraction of $q \,{\in}\, \{0.05,0.2\}$, altitude of $h\,{=}\,50\,\text{m}$, expansion ratio of $\kappa\,{=}\,3$ (i.e., $\Delta_\mathsf{e}\,{=}\,3 \Delta_\mathsf{u}$, $L_\mathsf{e}\,{=}\,3 L_\mathsf{u}$).}
\label{fig:rate_vs_Ptx_del5_h50m_rho3}
\vspace{0.0in}
\end{figure}

We investigate the impact of UAV-BS operation altitude by considering $h \,{=}\, 50\,\text{m}$, and produce the respective sum secrecy rate results in Fig.~\ref{fig:rate_vs_angle_del5_h50m_rho3} keeping all the other simulation parameters of Fig.~\ref{fig:rate_vs_angle_del5_h20m_rho3} the same. We observe that the sum secrecy rates deteriorate along with increasing altitude for the same area fraction $q$, as expected. Note that the effective channel gains of the NOMA users degrade with increasing altitude, thereby requiring larger protected zone to achieve the same secrecy rate performance. We also observe that the respective sum secrecy rates deteriorate rapidly as $\Delta_\mathsf{p}$ gets away from its optimal value, for which the corresponding decay characteristic at low altitude of $h \,{=}\, 20\,\text{m}$ is relatively slower, as presented in Fig.~\ref{fig:rate_vs_angle_del5_h20m_rho3}. We therefore conclude that \textit{the penalty associated with employing non-optimal protected zone shape (i.e., non-optimal $\Delta_\mathsf{p}$) becomes more significant along with increasing altitude}.  

We also present the secrecy-rate performance for $h \,{=}\, 50\,\text{m}$ along with varying transmit power in Fig.~\ref{fig:rate_vs_Ptx_del5_h50m_rho3} assuming $q \,{=}\, \{0.05,0.2\}$ and keeping all the other simulation parameters the same. We observe that the sum secrecy rates follow a very similar variation characteristic along with increasing transmit power as compared to those for $h \,{=}\, 20\,\text{m}$ depicted in Fig.~\ref{fig:Ptx_del5_h20m_rho3}\subref{fig:rate_vs_Ptx_del5_h20m_rho3}. In particular, the sum secrecy rate for the hybrid NOMA/SUT scheme with $q \,{=}\, 0.2$ does not exceed the target secrecy rate of the strong user (i.e, $4$ BPCU) for any transmit power. This result implies that the weak user cannot produce enough secrecy rate to meet its target secrecy rate at $h \,{=}\, 50\,\text{m}$ (as opposed to the corresponding low-altitude results in Fig.~\ref{fig:Ptx_del5_h20m_rho3}\subref{fig:rate_vs_Ptx_del5_h20m_rho3}). We also note that the secrecy-rate performance superiority of the hybrid NOMA/SUT schemes over OMA for this relatively higher altitude is much greater than the one for lower altitude of $h \,{=}\, 20\,\text{m}$ in Fig.~\ref{fig:Ptx_del5_h20m_rho3}\subref{fig:rate_vs_Ptx_del5_h20m_rho3}.

\begin{figure}[!t]
\centering
\subfloat[$h \,{=}\, 20\,\text{m}$] {\includegraphics[width=0.51\textwidth]{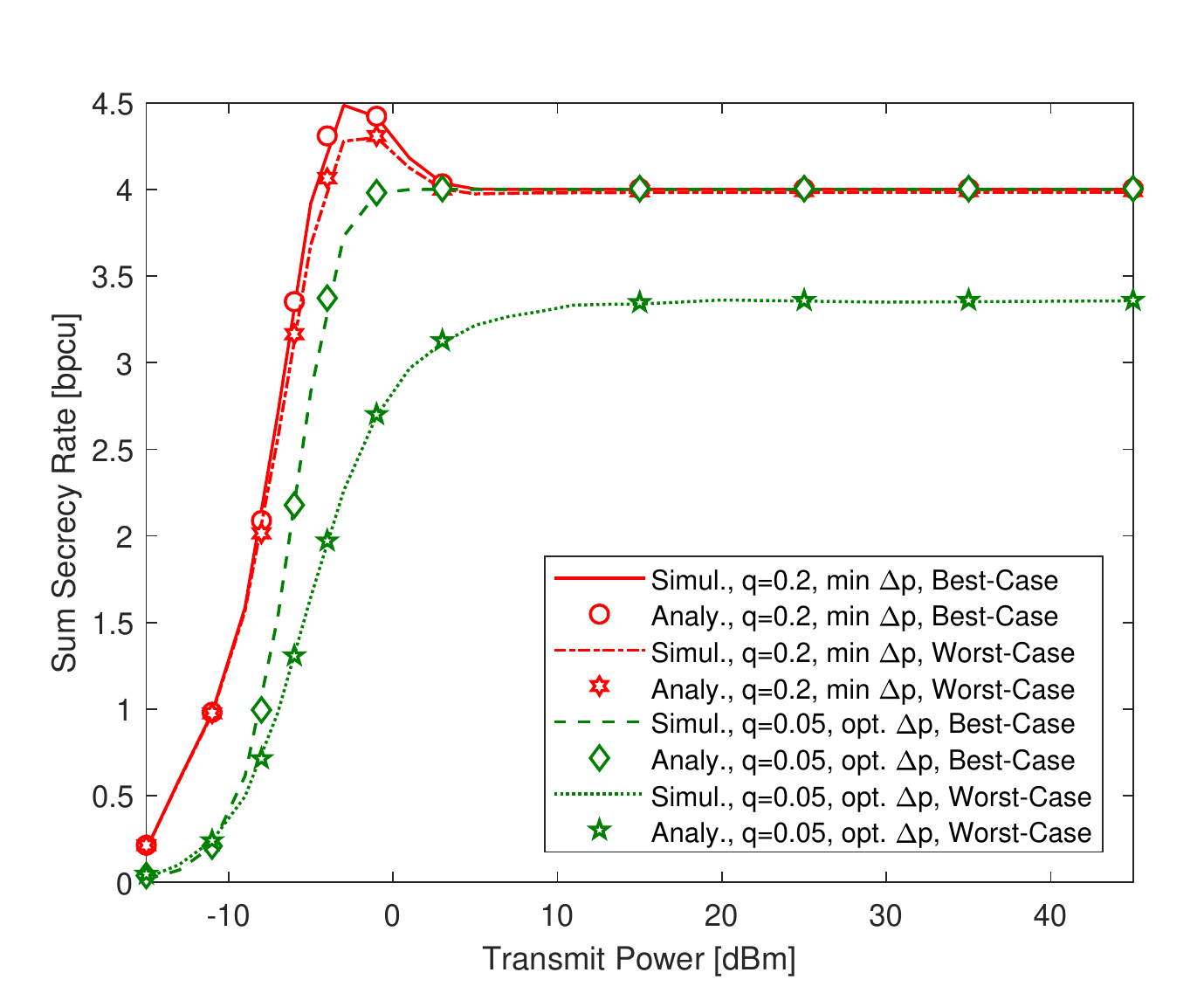}
\label{fig:rate_vs_Ptx_del5_h20m_rho3_bestcase}}\\
\vspace{0.0in}
\subfloat[$h \,{=}\, 50\,\text{m}$] {\includegraphics[width=0.5\textwidth]{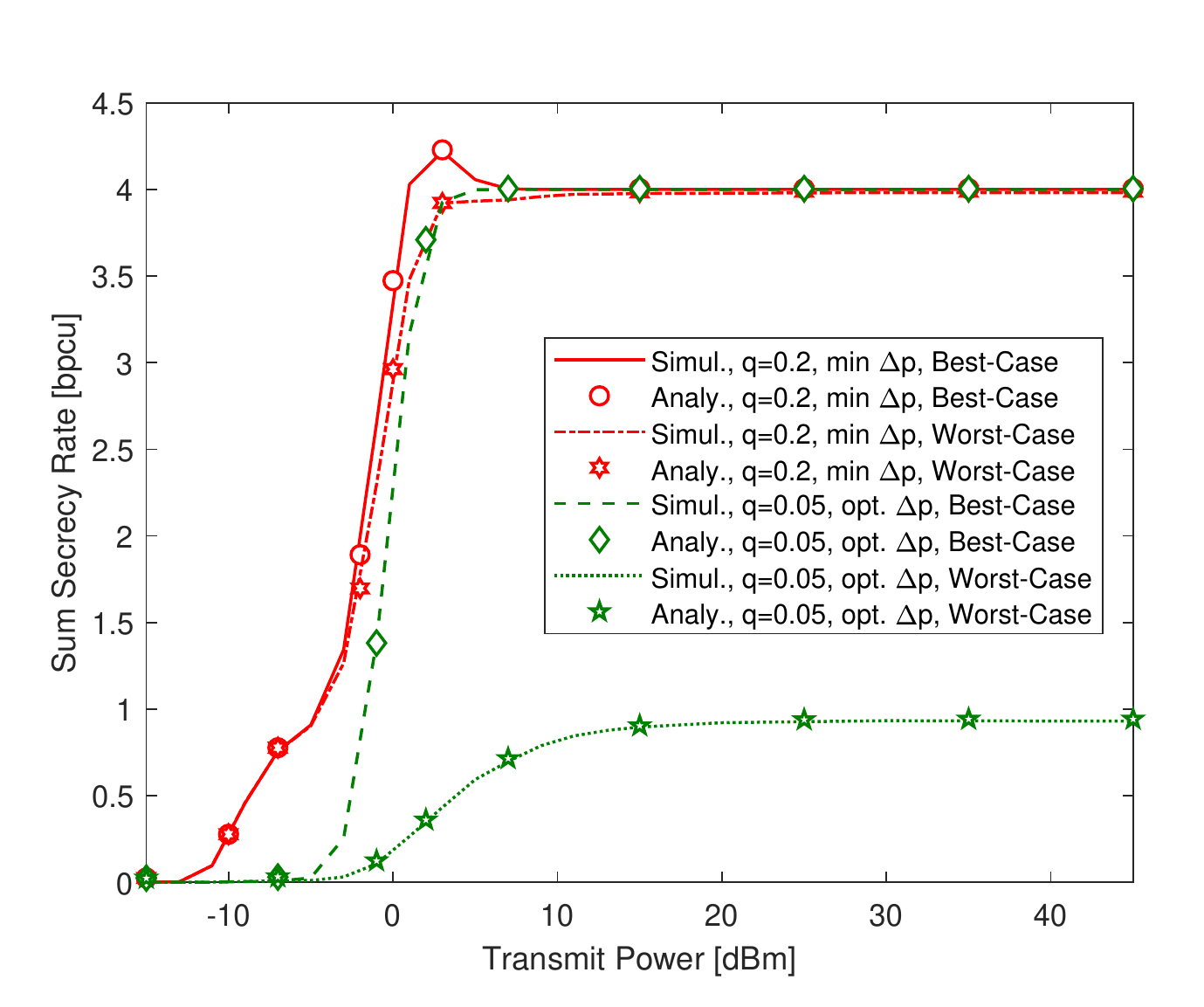}
\label{fig:rate_vs_Ptx_del5_h50m_rho3_bestcase}}
\caption{Sum secrecy rate vs. transmit power of hybrid NOMA/SUT scheme for worst- and best-case Eve decoding, assuming area fraction of $q \,{\in}\, \{0.05,0.2\}$, expansion ratio of $\kappa\,{=}\,3$ (i.e., $\Delta_\mathsf{e}\,{=}\,3 \Delta_\mathsf{u}$, $L_\mathsf{e}\,{=}\,3 L_\mathsf{u}$).}
\label{fig:rate_del5_h20m_rho2_bestcase}
\vspace{0.0in}
\end{figure}

We finally consider the impact of the decoding capability of the most detrimental Eve by taking into account the best-case condition, which assumes no multiuser decoding (i.e., no SIC) at the Eve side, as discussed in Section~\ref{subsec:decoding_performance_eve}. In particular, we depict the secrecy-rate performance of the hybrid NOMA/SUT scheme in Fig.~\ref{fig:rate_del5_h20m_rho2_bestcase} along with varying transmit power for worst- and best-case conditions at $h \,{=}\, \{20,50\}\,\text{m}$ while keeping all the other simulation parameters of Fig.~\ref{fig:rate_vs_angle_del5_h20m_rho3} the same. We observe that the sum secrecy rate of $q \,{=}\, 0.05$ catches that of $q \,{=}\, 0.2$ for the best-case condition, and saturates roughly for $\mathsf{P}_\mathsf{tx} \,{\geq}\, 5\,\text{dBm}$ irrespective of the UAV-BS altitude, as discussed in Remark~\ref{remark:eve_decoding_capability}. This result implies that as the most detrimental Eve loses its multiuser decoding capability (i.e., the scenario turns out to be the best-case condition), which might be due to many reasons including lack of information on decoding order or power allocation of the legitimate users, or error propagation in SIC, we do not need to have large protected zones (i.e., large $q$) at the high transmit-power regime. We also infer from the peak associated with $q \,{=}\, 0.2$ around $\mathsf{P}_\mathsf{tx} \,{=}\, 3\,\text{dBm}$ in Fig.~\ref{fig:rate_del5_h20m_rho2_bestcase}\subref{fig:rate_vs_Ptx_del5_h50m_rho3_bestcase} that the weak NOMA user can meet its target secrecy rate for the best-case condition at $h \,{=}\, 50\,\text{m}$, which never happens for the worst-case condition. We also note the huge improvement of $3$ BPCU in steady-state secrecy-rate performance as a result of the best-case condition (as compared to the worst-case condition).

\section{Conclusion}
\label{sec:conclusion}

In this work, we consider the sum-secrecy performance of NOMA-based transmission in a UAV-assisted mmWave network. The secrecy rates are improved through carefully designing protected zone outside the user region, which is cleared from the presence of any malicious receiver. We show that the optimal shape of the protected zone reflects a compromise between protecting angle and distance dimension, thanks to the mmWave propagation characteristics. We also show that the scenario-dependent practical values of protected zone area and transmit power are sufficient for the proposed strategy to achieve desired secrecy rates.    

\appendices

\section{Proof of Theorem~\ref{theorem:outage_strong}}
\label{appendix:outage_strong}

In this appendix, we derive the outage probability expressions for the strong NOMA and SUT users, in sequence, provided in Theorem~\ref{theorem:outage_strong}. To derive the secrecy outage probability for the strong NOMA user (i.e., $j$-th one), we elaborate \eqref{eq:outage_noma_4} considering the worst- and best-case conditions for the decoding capability of Eves separately. Employing the SINR expressions of \eqref{eq:sinr_user} and \eqref{eq:sinr_eve_worstcase}, and rate expressions of \eqref{eq:rate_noma} and \eqref{eq:rate_eve}, the desired secrecy outage probability in \eqref{eq:outage_noma_4} for the worst-case condition is given as 
\begin{align}
\mathsf{P}_{j\mid \Phi}^\mathsf{o} &= \mathsf{Pr} \bigg\lbrace \log \left( 1 + \frac{\mathsf{P}_\mathsf{tx}}{N_0} |\textbf{h}_{j}^{\rm H}\textbf{w}|^2 \beta_{j}^2 \right) \nonumber \\ 
& \hspace{0.8in} \leq \log \left( 1 + \frac{\mathsf{P}_\mathsf{tx}}{N_0} g_\mathsf{E} \beta_{j}^2  \right) + \overline{\mathsf{R}}_j \mid \Phi \bigg\rbrace , \label{eq:outage_noma_strong_1} \\
& = \mathsf{Pr} \bigg\lbrace |\textbf{h}_{j}^{\rm H}\textbf{w}|^2 \leq \frac{2^{\overline{\mathsf{R}}_j}{-}1}{ \frac{\mathsf{P}_\mathsf{tx}}{N_0} \beta_{j}^2}+2^{\overline{\mathsf{R}}_j}g_\mathsf{E}\mid \Phi  \bigg\rbrace . \label{eq:outage_noma_strong_2}
\end{align} 
Employing the PDF of the $j$-th user effective channel gain and that of the most detrimental Eve, \eqref{eq:outage_noma_strong_2} can be computed using \eqref{eq:outage_strong_sut} with the help of \eqref{eq:upper_limit_worst}.

For the best-case condition of decoding strong NOMA user, we employ the respective SINR expression in \eqref{eq:sinr_eve_bestcase}, and follow the same strategy of the worst-case condition which yields
\begin{align}
\mathsf{P}_{j\mid \Phi}^\mathsf{o} &= \mathsf{Pr} \bigg\lbrace \log \left( 1 + \frac{\mathsf{P}_\mathsf{tx}}{N_0}|\textbf{h}_{j}^{\rm H}\textbf{w}|^2 \beta_{j}^2 \right) \nonumber \\
& \hspace{0.3in} \leq \log \left( 1 +
\frac{\mathsf{P}_\mathsf{tx} \, g_\mathsf{E} \beta_{j}^2}{\mathsf{P}_\mathsf{tx} \, g_\mathsf{E} \beta_{i}^2 + N_0} \right) + \overline{\mathsf{R}}_j \mid \Phi \bigg\rbrace , \label{eq:outage_noma_strong_4} \\
& \hspace{-0.2in} =  \mathsf{Pr} \bigg\lbrace |\textbf{h}_{j}^{\rm H}\textbf{w}|^2 < \frac{2^{\overline{\mathsf{R}}_j}{-}1}{ \frac{\mathsf{P}_\mathsf{tx}}{N_0} \beta_{j}^2}+2^{\overline{\mathsf{R}}_j} \frac{g_\mathsf{E}}{1 +\frac{\mathsf{P}_\mathsf{tx}}{N_0} g_\mathsf{E} \beta_{i}^2} \mid \Phi \bigg\rbrace, \label{eq:outage_noma_strong_5}
\end{align}
which can be computed similarly by \eqref{eq:outage_strong_sut} and \eqref{eq:upper_limit_best}.

We finally consider the outage probability of the SUT user, which is very similar to that of the strong NOMA user. In particular, considering the SINR expressions of \eqref{eq:sinr_user} and \eqref{eq:sinr_eve_worstcase} (or equivalently \eqref{eq:sinr_eve_bestcase}) for single user only, and rate expressions of \eqref{eq:rate_noma} and \eqref{eq:rate_eve}, we express the outage probability in \eqref{eq:outage_sut_1} as follows 
\begin{align}
\mathsf{P}_{j\mid \Phi}^\mathsf{o} &= \mathsf{Pr} \bigg\lbrace \log \left( 1 + \frac{\mathsf{P}_\mathsf{tx}}{N_0} |\textbf{h}_{j}^{\rm H}\textbf{w}|^2  \right) \nonumber \\
& \hspace{0.8in} \leq \log \left( 1 + \frac{\mathsf{P}_\mathsf{tx}}{N_0} g_\mathsf{E} \right) + \overline{\mathsf{R}}_j \mid \Phi \bigg\rbrace , \label{eq:outage_sut_2} \\
& = \mathsf{Pr} \left\lbrace |\textbf{h}_{j}^{\rm H}\textbf{w}|^2 \leq \frac{2^{\overline{\mathsf{R}}_j}{-}1}{ \frac{\mathsf{P}_\mathsf{tx}}{N_0} }+2^{\overline{\mathsf{R}}_j}g_\mathsf{E}\mid \Phi  \right\rbrace , \label{eq:outage_sut_3}
\end{align} 
which can be calculated similarly by \eqref{eq:outage_strong_sut} along with the upper limit.
where $\delta^\mathsf{max}_j(y) \,{=}\, \frac{2^{\overline{\mathsf{R}}_j}{-}1}{ \frac{\mathsf{P}_\mathsf{tx}}{N_0} }{+}2^{\overline{\mathsf{R}}_j}y$.

\section{Proof of Theorem~\ref{theorem:outage_weak}}
\label{appendix:outage_weak}

In this appendix, we derive the outage probability expression for the weak NOMA user (i.e., $i$-th one) given in Theorem~\ref{theorem:outage_weak}. We follow the same strategy of the outage probability derivation for the strong user while carrying out the desired derivation. Note that the SINR expressions in \eqref{eq:sinr_eve_worstcase} and \eqref{eq:sinr_eve_bestcase} corresponding to the worst- and best-case conditions become equal for the weak user. This observation is because of the fact that the most detrimental Eve \textit{always} decodes the weak NOMA user message in the presence of the strong user's message, and, therefore, does not apply SIC. 

The desired outage probability for both the worst- and best-case condition is given as
\begin{align} 
\mathsf{P}_{i\mid \Phi}^\mathsf{o} &= \mathsf{Pr} \bigg\lbrace \log \left( 1 + \frac{\mathsf{P}_\mathsf{tx}|\textbf{h}_{i}^{\rm H}\textbf{w}|^2 \beta_{i}^2}{ \mathsf{P}_\mathsf{tx}|\textbf{h}_{i}^{\rm H}\textbf{w}|^2 \beta_{j}^2 + N_0} \right) \nonumber \\
& \hspace{0.5in} \leq \log \left( 1 + \frac{\mathsf{P}_\mathsf{tx}g_\mathsf{E} \beta_{i}^2}{ \mathsf{P}_\mathsf{tx}g_\mathsf{E} \beta_{j}^2 + N_0} \right) + \overline{\mathsf{R}}_i \mid \Phi \bigg\rbrace , \label{eq:outage_noma_weak_1} \\
&= \mathsf{Pr} \left\lbrace \varepsilon_1  \, |\textbf{h}_{i}^{\rm H}\textbf{w}|^2 < \varepsilon_2  \mid \Phi  \right\rbrace \label{eq:outage_noma_weak_2} ,
\end{align}
where
\begin{align}
    \varepsilon_1 &= \frac{\mathsf{P}_\mathsf{tx}}{N_0} \left( (1+\frac{\mathsf{P}_\mathsf{tx}}{N_0}\beta_{j}^2 g_\mathsf{E}) - 2^{\overline{\mathsf{R}}_i}\beta_{j}^2(1+\frac{\mathsf{P}_\mathsf{tx}}{N_0} g_\mathsf{E} )  \right) , 
    \label{eq:outer_limit_den} \\
    \varepsilon_2 &= 2^{\overline{\mathsf{R}}_i}(1+\frac{\mathsf{P}_\mathsf{tx}}{N_0} g_\mathsf{E}) - (1+\frac{\mathsf{P}_\mathsf{tx}}{N_0}\beta_{j}^2 g_\mathsf{E}). \label{eq:outer_limit_nom} 
\end{align}
Based on $\varepsilon_1$ in \eqref{eq:outer_limit_den} being negative or positive, the desired outage probability in \eqref{eq:outage_noma_weak_2} is further elaborated as follows
\begin{subnumcases}{\mathsf{P}_{i\mid \Phi}^\mathsf{o} =}
\mathsf{Pr} \left\lbrace |\textbf{h}_{i}^{\rm H}\textbf{w}|^2 < \varepsilon_2/\varepsilon_1 \mid \Phi  \right\rbrace , & $\varepsilon_1 > 0$, \label{eq:outage_noma_weak_3} \\
\mathsf{Pr} \left\lbrace |\textbf{h}_{i}^{\rm H}\textbf{w}|^2 > \varepsilon_2/\varepsilon_1 \mid \Phi  \right\rbrace , & $\varepsilon_1 < 0$, \label{eq:outage_noma_weak_4}
\end{subnumcases}
where $\varepsilon_1 > 0$ is equivalent to 
\begin{align}
    g_\mathsf{E} < \varrho = \frac{ 1 - 2^{\overline{\mathsf{R}}_i} \beta_j^2 }{ \frac{\mathsf{P}_\mathsf{tx}}{N_0} \beta_j^2 \left( 2^{\overline{\mathsf{R}}_i} - 1 \right) }. \label{app:outer_limit}
\end{align}
The desired outage probability can then be computed using \eqref{eq:outage_weak} by the help of \eqref{eq:threshold_weak} and \eqref{eq:outer_limit}.

\section{Proof of Theorem~\ref{theorem:cdf_unordered_eve}}
\label{appendix:cdf_unordered_eve}

In this appendix, we derive the unordered CDF of Eves after the protected zone of either type specified in Section~\ref{subsec:protected_zone} is subtracted from the original Eve region, which is referred to as unprotected-Eve region. The desired CDF is defined for an arbitrary Eve with the index $k$ as
\begin{align}\label{eq:cdf_eve_1}
F_\mathsf{E}(z) &= \mathsf{Pr} \left\lbrace |\textbf{g}_k^{\rm H} \textbf{w}|^2 \leq z \right\rbrace \\
&= \int_{\overline{\theta}{-}\Delta_\mathsf{e}}^{\overline{\theta}{+}\Delta_\mathsf{e}} \!\! \int_{L_\mathsf{i}}^{L_\mathsf{e}} \mathsf{Pr} \left\lbrace |\textbf{g}_k^{\rm H} \textbf{w}|^2 \,{\leq}\, z \,{\mid}\, \theta_k {=}\, \theta, d_k {=}\, r \right\rbrace \nonumber \\
& \hspace{1.5in} \times f_\mathsf{EP}(\theta,r) \dd r \dd \theta,
\end{align}
where $f_\mathsf{EP}(\theta,r)$ is the PDF of the location (i.e., joint angle-distance) distribution for Eves in the unprotected-Eve region. Note that $f_\mathsf{EP}(\theta,r)$ is seemingly a function of the specific shape of the protected zone, and, hence, depends on particular choice of $(\Delta_\mathsf{p},L_\mathsf{p})$. Considering the effective channel gain definition of \eqref{eq:effective_channel_gain}, the CDF in \eqref{eq:cdf_eve_1} can be further elaborated as follows
\begin{align}\label{eq:cdf_eve_2}
&F_\mathsf{E}(z) \,{=} \int_{\overline{\theta}{-}\Delta_\mathsf{e}}^{\overline{\theta}{+}\Delta_\mathsf{e}} \!\!\! 
\int_{L_\mathsf{i}}^{L_\mathsf{e}} \!\! \mathsf{Pr} \bigg\lbrace \frac{|\alpha_k|^2}{\mathsf{PL}\left(\sqrt{r^2 + h^2}\right)} \mathsf{F}_M \left(\overline{\theta}, \theta_k \right) \leq z \bigg\rbrace \nonumber \\
& \hspace{2in} \times f_\mathsf{EP}(\theta,r) \dd r \dd \theta .
\end{align}
Recalling that $|\alpha_k|^2$ is exponentially distributed due to the definition of small-scale fading $\alpha_k$ being complex Gaussian, the desired CDF in \eqref{eq:cdf_eve_2} becomes \cite{Yapici2019AngFee}
\begin{align}\label{eq:cdf_eve_3}
F_\mathsf{E}(z) = \int_{\overline{\theta}{-}\Delta_\mathsf{e}}^{\overline{\theta}{+}\Delta_\mathsf{e}} \int_{L_\mathsf{i}}^{L_\mathsf{e}}  \Bigg( 1 \,{-}\, e^{ \frac{-z{\mathsf{PL} \big(\sqrt{r^2 + h^2}\big)}}{ {\mathsf{F}_M \left(\overline{\theta}, \theta_k \right)}} }  \Bigg) f_\mathsf{EP}(\theta,r) \dd r \dd \theta.
\end{align}

Considering an arbitrary angular sector within the unprotected-Eve region of Fig.~\ref{fig:protected_zone} with a constant minimum radius of $L$ and angular width of $\Delta$, the distance and angle of Eves are statistically independent with the marginal PDF of the distance given as \cite{Yapici2019AngFee}
\begin{align}
    f_d(r) = \frac{2r}{L_\mathsf{e}^2-L^2}.
\end{align}
Since the angle is uniformly distributed, the conditional PDF of the location is given as
\begin{align} \label{eq:pdf_location_conditional}
f_\mathsf{EP}(\theta,r | \mathcal{S}_\mathcal{A}) = \frac{2r}{L_\mathsf{e}^2-L^2} \times\frac{1}{\Delta} = \frac{r}{\mathsf{A}} ,
\end{align}
where $\mathcal{S}_\mathcal{A}$ stands for the set of all angle-distance pairs within the specified angular sector, and $\mathsf{A}$ is the respective area (of the angular sector). The unconditional PDF of the location can be obtained using \eqref{eq:pdf_location_conditional} as follows
\begin{align}
f_\mathsf{EP}(\theta,r) &= \int_{\mathcal{S}_\mathcal{A}} f_\mathsf{EP}(\theta,r | \mathcal{S}_\mathcal{A} = y) \,  f_{\mathcal{S}_\mathcal{A}}( y) \, \dd y \label{eq:pdf_location_unconditional_1} \\
&= \int_{\mathcal{S}_\mathcal{A}} \frac{r}{\mathsf{A}} \times \frac{\mathsf{A}}{\mathsf{A}_\mathsf{e}-\mathsf{A}_\mathsf{p}} \, \dd y =  \frac{r}{\mathsf{A}_\mathsf{e}-\mathsf{A}_\mathsf{p}}, \label{eq:pdf_location_unconditional_2}
\end{align}
which uses the fact that any angle-distance pair is present in the unprotected-Eve region with the probability of $\frac{\mathsf{A}}{\mathsf{A}_\mathsf{EP}}$, thanks to the assumption of Eve deployment being homogeneous. Note also that $f_\mathsf{EP}(\theta,r)$ takes nonzero values only when the angle-distance pair is in the unprotected-Eve region, i.e., $(\theta,r) \,{\in}\, \mathcal{S}_\mathsf{EP}$. Since the integration in \eqref{eq:cdf_eve_3} considers also the angle-distance pairs such that $(\theta,r) \,{\not\in}\, \mathcal{S}_\mathsf{EP}$, we describe the possible $(\theta,r)$ pairs by the indicator function in \eqref{eq:indicator}, which readily yields \eqref{eq:cdf_unordered_eves} using \eqref{eq:cdf_eve_3} and \eqref{eq:pdf_location_unconditional_2}. \IEEEQEDhere

\bibliographystyle{IEEEtran}
\bibliography{references}

\begin{thebibliography}{10}
\providecommand{\url}[1]{#1}
\csname url@samestyle\endcsname
\providecommand{\newblock}{\relax}
\providecommand{\bibinfo}[2]{#2}
\providecommand{\BIBentrySTDinterwordspacing}{\spaceskip=0pt\relax}
\providecommand{\BIBentryALTinterwordstretchfactor}{4}
\providecommand{\BIBentryALTinterwordspacing}{\spaceskip=\fontdimen2\font plus
\BIBentryALTinterwordstretchfactor\fontdimen3\font minus
  \fontdimen4\font\relax}
\providecommand{\BIBforeignlanguage}[2]{{%
\expandafter\ifx\csname l@#1\endcsname\relax
\typeout{** WARNING: IEEEtran.bst: No hyphenation pattern has been}%
\typeout{** loaded for the language `#1'. Using the pattern for}%
\typeout{** the default language instead.}%
\else
\language=\csname l@#1\endcsname
\fi
#2}}
\providecommand{\BIBdecl}{\relax}
\BIBdecl

\bibitem{Cisco2019visnet}
\BIBentryALTinterwordspacing
Cisco. (2019, Jun.) {Cisco Visual Networking Index: Forecast and Trends,
  2017--2022}. [Online]. Available:
  \url{https://www.cisco.com/c/en/us/solutions/service-provider/visual-networking-index-vni}
\BIBentrySTDinterwordspacing

\bibitem{Xia2019mmW}
L.~{Zhu}, Z.~{Xiao}, X.~{Xia}, and D.~{Oliver Wu}, ``Millimeter-wave
  communications with non-orthogonal multiple access for {B5G/6G},'' \emph{IEEE
  Access}, vol.~7, pp. 116\,123--116\,132, Aug. 2019.

\bibitem{Zhang2019uavCom}
B.~{Li}, Z.~{Fei}, and Y.~{Zhang}, ``{UAV} communications for {5G} and beyond:
  {Recent} advances and future trends,'' \emph{IEEE Internet Things J.},
  vol.~6, no.~2, pp. 2241--2263, Apr. 2019.

\bibitem{Arslan2019PHYSECSur}
J.~M. Hamamreh, H.~M. Furqan, and H.~Arslan, ``Classifications and applications
  of physical layer security techniques for confidentiality: {A} comprehensive
  survey,'' \emph{IEEE Commun. Surveys Tuts.}, vol.~21, no.~2, pp. 1773--1828,
  2nd Quar. 2019.

\bibitem{Swami2013PHYLay}
N.~Romero-Zurita, D.~McLernon, M.~Ghogho, and A.~Swami, ``{PHY} layer security
  based on protected zone and artificial noise,'' \emph{IEEE Sig. Process.
  Lett.}, vol.~20, no.~5, pp. 487--490, May 2013.

\bibitem{Quek2014EnhSec}
S.~H. {Chae}, W.~{Choi}, J.~H. {Lee}, and T.~Q.~S. {Quek}, ``Enhanced secrecy
  in stochastic wireless networks: {Artificial} noise with secrecy protected
  zone,'' \emph{IEEE Trans. Inf. Forensics Security}, vol.~9, no.~10, pp.
  1617--1628, Oct. 2014.

\bibitem{Ding2016OnErg}
W.~{Liu}, Z.~{Ding}, T.~{Ratnarajah}, and J.~{Xue}, ``On ergodic secrecy
  capacity of random wireless networks with protected zones,'' \emph{IEEE
  Trans. Vehic. Technol.}, vol.~65, no.~8, pp. 6146--6158, Aug. 2016.

\bibitem{Negi2008GuaSec}
S.~{Goel} and R.~{Negi}, ``Guaranteeing secrecy using artificial noise,''
  \emph{IEEE Trans. Wireless Commun.}, vol.~7, no.~6, pp. 2180--2189, Jun.
  2008.

\bibitem{Yin21017SecTra}
Y.~{Ju}, H.~{Wang}, T.~{Zheng}, and Q.~{Yin}, ``Secure transmissions in
  millimeter wave systems,'' \emph{IEEE Trans. Commun.}, vol.~65, no.~5, pp.
  2114--2127, May 2017.

\bibitem{Jiang2019UAV}
H.~{Wang}, X.~{Zhang}, and J.~{Jiang}, ``{UAV}-involved wireless physical-layer
  secure communications: Overview and research directions,'' \emph{IEEE
  Wireless Commun.}, vol.~26, no.~5, pp. 32--39, Oct. 2019.

\bibitem{Ng2019PhyLay}
X.~{Sun}, D.~W.~K. {Ng}, Z.~{Ding}, Y.~{Xu}, and Z.~{Zhong}, ``Physical layer
  security in {UAV} systems: Challenges and opportunities,'' \emph{IEEE
  Wireless Commun.}, vol.~26, no.~5, pp. 40--47, Oct. 2019.

\bibitem{Zeng2019PhyLay}
N.~{Wang}, P.~{Wang}, A.~{Alipour-Fanid}, L.~{Jiao}, and K.~{Zeng},
  ``Physical-layer security of {5G} wireless networks for {IoT}: Challenges and
  opportunities,'' \emph{IEEE Internet Things J.}, vol.~6, no.~5, pp.
  8169--8181, Oct. 2019.

\bibitem{Zhang2020UAV}
H.-M. {Wang} and X.~{Zhang}, ``{UAV} secure downlink {NOMA} transmissions: {A}
  secure users oriented perspective,'' \emph{IEEE Trans. Commun.}, vol.~68,
  no.~9, pp. 5732--5746, Sep. 2020.

\bibitem{Sun2019SecTra}
X.~{Sun}, W.~{Yang}, Y.~{Cai}, Z.~{Xiang}, and X.~{Tang}, ``Secure
  transmissions in millimeter wave {SWIPT UAV}-based relay networks,''
  \emph{IEEE Wireless Commun. Lett.}, vol.~8, no.~3, pp. 785--788, Jun. 2019.

\bibitem{Sun2019PhyLay}
X.~{Sun}, W.~{Yang}, Y.~{Cai}, R.~{Ma}, and L.~{Tao}, ``Physical layer security
  in millimeter wave {SWIPT UAV}-based relay networks,'' \emph{IEEE Access},
  vol.~7, pp. 35\,851--35\,862, Mar. 2019.

\bibitem{Sun2019SecCom}
X.~{Sun}, W.~{Yang}, and Y.~{Cai}, ``Secure communication in {NOMA} assisted
  millimeter wave {SWIPT UAV} networks,'' \emph{IEEE Internet Things J.},
  vol.~7, no.~3, pp. 1884--1897, Mar. 2020.

\bibitem{Cai2020SecmmW}
X.~{Sun}, W.~{Yang}, Y.~{Cai}, and M.~{Wang}, ``Secure {MmWave} {UAV}-enabled
  {SWIPT} networks based on random frequency diverse arrays,'' \emph{IEEE
  Internet Things J.}, 2020, {Early Access}.

\bibitem{Shi2019SecmmW}
R.~{Ma}, W.~{Yang}, Y.~{Zhang}, J.~{Liu}, and H.~{Shi}, ``Secure {mmWave}
  communication using {UAV}-enabled relay and cooperative jammer,'' \emph{IEEE
  Access}, vol.~7, pp. 119\,729--119\,741, 2019.

\bibitem{Dai2018ReiLea}
L.~{Xiao}, Y.~{Li}, C.~{Dai}, H.~{Dai}, and H.~V. {Poor}, ``Reinforcement
  learning-based {NOMA} power allocation in the presence of smart jamming,''
  \emph{IEEE Trans. Vehic. Technol.}, vol.~67, no.~4, pp. 3377--3389, Apr.
  2018.

\bibitem{Yapici2018EnhPhy}
N.~Rupasinghe, Y.~Yap{\i}c{\i}, I.~G\"{u}ven\c{c}, H.~Dai, and A.~Bhuyan,
  ``Enhancing physical layer security for {NOMA} transmission in {mmWave} drone
  networks,'' in \emph{Proc. Asilomar Conf. Signals, Syst., and Comput.},
  Pacific Grove, California, Oct. 2018, pp. 729--733.

\bibitem{Ding17PoorRandBeamforming}
Z.~Ding, P.~Fan, and H.~V. Poor, ``Random beamforming in millimeter-wave {NOMA}
  networks,'' \emph{IEEE Access}, vol.~5, pp. 7667--7681, Feb. 2017.

\bibitem{Hanzo2017EnhPhy}
Y.~Liu, Z.~Qin, M.~Elkashlan, Y.~Gao, and L.~Hanzo, ``Enhancing the physical
  layer security of non-orthogonal multiple access in large-scale networks,''
  \emph{IEEE Trans. Wireless Commun.}, vol.~16, no.~3, pp. 1656--1672, Mar.
  2017.

\bibitem{Ding2016SecSum}
Y.~{Zhang}, H.~{Wang}, Q.~{Yang}, and Z.~{Ding}, ``Secrecy sum rate
  maximization in non-orthogonal multiple access,'' \emph{IEEE Commun. Lett.},
  vol.~20, no.~5, pp. 930--933, May 2016.

\bibitem{Yuan2019BeaDes}
Y.~{Feng}, S.~{Yan}, Z.~{Yang}, N.~{Yang}, and J.~{Yuan}, ``Beamforming design
  and power allocation for secure transmission with {NOMA},'' \emph{IEEE Trans.
  Wireless Commun.}, vol.~18, no.~5, pp. 2639--2651, May 2019.

\bibitem{Lau2017OnDes}
B.~{He}, A.~{Liu}, N.~{Yang}, and V.~K.~N. {Lau}, ``On the design of secure
  non-orthogonal multiple access systems,'' \emph{IEEE J. Sel. Areas Commun.},
  vol.~35, no.~10, pp. 2196--2206, Oct. 2017.

\bibitem{Qin2017SecSum}
M.~{Tian}, Q.~{Zhang}, S.~{Zhao}, Q.~{Li}, and J.~{Qin}, ``Secrecy sum rate
  optimization for downlink {MIMO} nonorthogonal multiple access systems,''
  \emph{IEEE Signal Process. Lett.}, vol.~24, no.~8, pp. 1113--1117, Aug. 2017.

\bibitem{Ding2019SecTra}
N.~{Zhao}, D.~{Li}, M.~{Liu}, Y.~{Cao}, Y.~{Chen}, Z.~{Ding}, and X.~{Wang},
  ``Secure transmission via joint precoding optimization for downlink {MISO
  NOMA},'' \emph{IEEE Trans. Vehic. Technol.}, vol.~68, no.~8, pp. 7603--7615,
  Aug. 2019.

\bibitem{Ng2018ExpInt}
X.~{Chen}, Z.~{Zhang}, C.~{Zhong}, D.~W.~K. {Ng}, and R.~{Jia}, ``Exploiting
  inter-user interference for secure massive non-orthogonal multiple access,''
  \emph{IEEE J. Sel. Areas Commun.}, vol.~36, no.~4, pp. 788--801, Apr. 2018.

\bibitem{Tsiftsis2019SecUse}
H.~{Wang}, X.~{Zhang}, Q.~{Yang}, and T.~A. {Tsiftsis}, ``Secure users oriented
  downlink {MISO NOMA},'' \emph{IEEE J. Sel. Topics Signal Process.}, vol.~13,
  no.~3, pp. 671--684, Jun. 2019.

\bibitem{David03Order_stat}
H.~A. David and H.~N. Nagaraja, \emph{Order Statistics}.\hskip 1em plus 0.5em
  minus 0.4em\relax Willey, New York, {Nf}, {USA}, 2003.

\bibitem{Yapici2019AngFee}
N.~Rupasinghe, Y.~Yap{\i}c{\i}, I.~G\"{u}ven\c{c}, M.~Ghosh, and Y.~Kakishima,
  ``Angular feedback for {mmWave NOMA} drone networks,'' \emph{{IEEE} J. Sel.
  Topics Signal Process.}, vol.~13, no.~3, pp. 628--643, Jun. 2019.

\bibitem{3GPP_TR38901}
{3GPP TR38.901}, ``Study on channel model for frequencies from $0.5$ to $100$
  {GHz} ({Release} $16$),'' 3rd {Generation Partnership Project (3GPP)}, Tech.
  Rep., Oct. 2019.

\end{thebibliography}
\end{document}